\documentclass[journal]{IEEEtran}
\ifCLASSINFOpdf
\else
\fi
\usepackage{cite}
\usepackage{graphicx} 
\usepackage{amsmath}
\usepackage{mathtools}
\usepackage{amssymb}
\usepackage{amsthm}
\usepackage{xcolor}
\usepackage{booktabs, tabularx}
\usepackage{fixdif}
\usepackage{empheq}
\usepackage{comment}
\usepackage{tikz}
\usetikzlibrary{arrows.meta,calc}
\usepackage{bm}
\usepackage[hidelinks]{hyperref}  
\usepackage{longtable}

\begin{document}
\title{Dynamic Modeling of Data-Center Power Delivery for Power System Resonance Analysis}
\author{Xingyu~Zhao,~\IEEEmembership{Student~Member,~IEEE,~}and~Junbo~Zhao,~\IEEEmembership{Senior~Member,~IEEE}
\thanks{This work is partially supported by the U.S. National Science Foundation under grant 2442160.}
\thanks{X. Zhao and J. Zhao are with Thayer School of Engineering at Dartmouth College, Hanover, NH 06269, USA. (e-mails: xingyu.zhao.th@dartmouth.edu, junbo.zhao@dartmouth.edu)}}
\markboth{Submitted to IEEE Transactions on Power Systems}%
{Shell \MakeLowercase{\textit{et al.}}: Bare Demo of IEEEtran.cls for IEEE Journals}
\maketitle

\begin{abstract}
The rapid proliferation of data centers is reshaping modern power system dynamics. Unlike legacy industrial loads, data centers have power-electronic interfaces whose multi-timescale dynamics can interact strongly with the grid, inducing oscillatory behavior. However, analytical models that are grid-integratable for revealing the underlying resonance mechanisms remain largely unexplored. To fill this research gap, this paper derives an explicit, component-informed dynamic model of data-center power-delivery chains, which preserves component-level fidelity and captures inter-stage control interactions. This model is formulated as a time-invariant representation in the positive-sequence domain, enabling seamless integration with the phasor (or RMS) domain power-system dynamic models. The analytical derivation reveals how realistic server-load fluctuations at specific frequencies can excite coupled control modes, thereby inducing oscillation amplification and propagation in power grids with heterogeneous dynamic resources, including synchronous machines and grid-forming/following inverters. Case studies on test systems with some realistic data center data demonstrate the effectiveness of the proposed solutions.
\end{abstract}
\IEEEpeerreviewmaketitle

\vspace{-0.3cm}
\section{Introduction}
The rapid expansion of AI- and cloud-enabled data centers induces a significant demand growth, with global data-center consumption projected to rise sharply through 2030 \cite{IEA_EnergyAI_2024}. Unlike conventional industrial loads, data centers are interfaced through cascaded power-electronic stages whose multi-timescale physical dynamics and control interaction reduce small-signal damping of local modes, and thus elevate resonance risk. Meanwhile, AI-driven, fast power fluctuations can act as sustained forcing inputs that excite these modes and lead to forced oscillations \cite{Valverde_AIFO_2025}. These risks and the need for mitigation are increasingly emphasized in industry guidance for emerging large loads \cite{NERC_RiskMitigation_EmergingLargeLoads_2026}.  However, today’s data-center interconnection studies often rely on either generic load model that obscures component-level root causes, or a detailed black-box electromagnetic-transient (EMT) model that provides limited transparency for mechanism interpretation and efficient damping design. Therefore, an explicit, component-informed model is needed to capture the critical multi-stage couplings and enable direct mechanism tracing via modal analysis. 

\subsection{Literature Review}
\subsubsection{Data-Center-Induced Oscillation}
The resonance oscillation risk introduced by grid-connected data centers has also drawn growing community attention. In \cite{Mishra_SEGAN_2025}, a $\sim$14.7--14.8~Hz oscillation event emerging from a real data center is observed, offering rare field evidence that data-center/grid interactions can manifest as grid-visible oscillatory phenomena. However, the underlying oscillatory mechanism remains unclear to the community. In~\cite{Valverde_AIFO_2025}, AI-workload power swings are modeled as sustained forcing disturbances that can excite weakly damped modes and thereby induce forced oscillations. The potential of AI load fluctuations to amplify both local and inter-area oscillation modes in large-scale systems is further shown in~\cite{KoZhu_arXiv_2025}.  Complementary to these system-level studies, an impedance-based stability framework is proposed in \cite{Sun_JPES_DC_Imp_I_2022,Sun_JPES_DC_Imp_II_2022} motivated by real data-center resonance incidents. These works reveal the intrinsic resonance risk arising from the converter-dominated power distribution system. However, a unified understanding of how intrinsic resonances propagate through the data-center power-delivery chain and produce grid-coupled oscillatory phenomena is still missing.

\subsubsection{Data-Center Dynamic Modeling}
In recent years, there has been growing community interest in developing dynamic models of data centers for transmission system interconnection and stability studies. In \cite{JimenezRuiz_Milano_DC_TS_2025}, a transmission-oriented data-center load model is proposed to study transient stability and fault-ride behavior. However, the propagation of grid-side power disturbances to the IT load is captured by a high-level aggregated representation, so the multi-stage interactions between individual converters and the grid are unmodeled. \cite{Gyang_TPWRS_DC_2025} develops a utility-interconnected data-center dynamic model that couples server power/heat with detailed cooling and air-handling system dynamics. However, the converter-dominated electrical dynamics remain largely simplified. Recently, EMT models that capture detailed power-electronic converters have been developed in~\cite{Sun_TPEL_DCEmu_2022,RossFollum_PNNL_DML_2025}. However, those waveform-domain models are not readily amenable to model-based small-signal analysis that relies on time-invariant, positive-sequence representations. Up to now, an explicit differential-algebraic model with component-level fidelity that can be seamlessly integrated into classical positive-sequence power-system oscillation analysis remains largely unexplored.

 \begin{figure*}[!tb]
     \centering
     \includegraphics[width=0.9\linewidth]{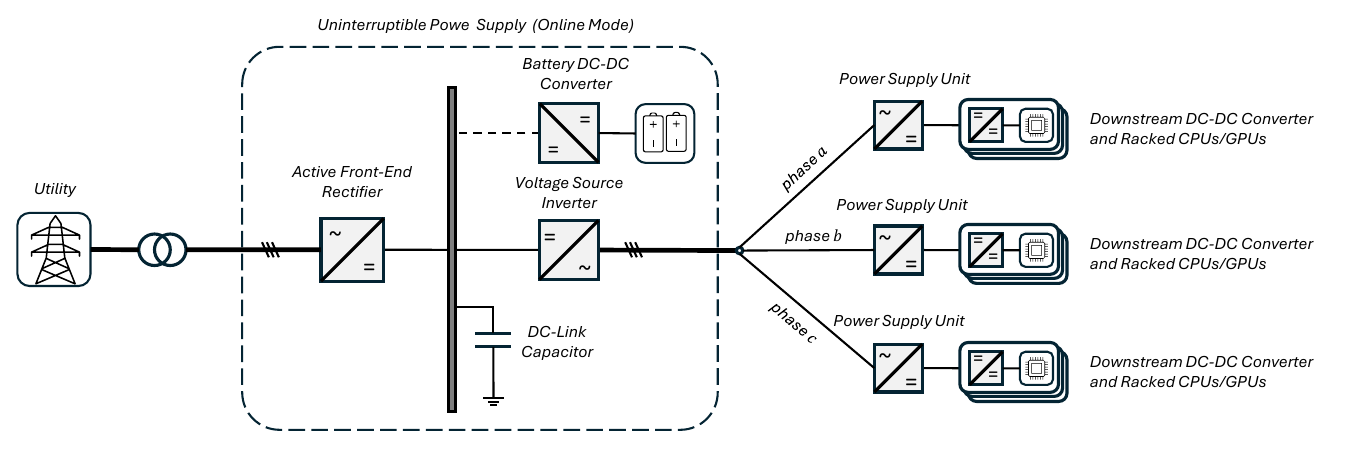}
     \caption{The modern data-center power-delivery system operating in online (double-conversion) mode}
     \label{fig:online}
 \end{figure*}

\subsection{Contributions} 
To fill these research gaps, this paper develops a component-informed model of the data-center power-delivery chain that can be integrated with the bulk power system for dynamic stability analysis. The major contributions of this paper are:
\begin{itemize}
\item \emph{Component-Informed Modeling:} We derive a composite dynamic model covering the full online power flow path, preserving key device physics and control structure while being suitable for power-system studies.
\item \emph{Grid Integration:} We propose a time-invariant reduced model in the positive-sequence domain, which maintains good accuracy compared with the full-order three-phase model while enabling seamless integration into standard phasor-domain simulation as well as small-signal stability analysis workflows.
\item \emph{Oscillation Analysis:} We examine how real-world server-load perturbations (especially those driven by AI training workload) can excite weakly-damped coupled-control modes along the data-center power path, amplifying oscillations and injecting them into power grids with heterogeneous dynamics (e.g., synchronous machines and inverter-based resources).
\end{itemize}

The remainder of this paper is organized as follows. Section~\ref{sec:dae} derives the dynamic model of the data-center power-delivery chain under the online operating mode. Section~\ref{sec:osc} presents the oscillation-analysis framework, including modal analysis, participation factors, and a power-oscillation amplification metric. Section~\ref{sec:case} reports case studies to illustrate intrinsic and grid-coupled oscillation mechanisms. Section~\ref{con} concludes the paper and discusses future work.

\section{Mathematical Modeling}
\label{sec:dae}
Fig.~\ref{fig:online} illustrates a typical data-center power-delivery system that operates in the online (double-conversion) mode of an uninterruptible power supply (UPS). Utility power is processed by an active front-end (AFE) rectifier, buffered by a DC-link capacitor,  and converted by a voltage-source inverter (VSI) to a regulated three-phase AC bus that feeds an array of single-phase rack-level power supply units (PSUs). Each PSU supplies downstream DC--DC converters and aggregated CPU/GPU loads. A battery storage system (BSS) interfaced by a DC--DC converter, is connected to the DC link and remains in standby (float-charge) mode during normal operation. \textbf{It is emphasized that our mathematical modeling focuses on the online operating mode}. Bypass operation and mode-transfer transients, typically associated with protection or maintenance and following a different power path, are outside the scope of this study. Battery-only operation is also not considered, as it corresponds to islanded conditions without grid connection.

In this section, we will start our derivations with detailed component-level differential-algebraic equations (DAEs) and merge them to a high-fidelity dynamic model of the complete data-center power-delivery chains.

\subsection{Active Front-End Rectifier}
\begin{figure}[!tb]
     \centering
     \includegraphics[width=0.8\linewidth]{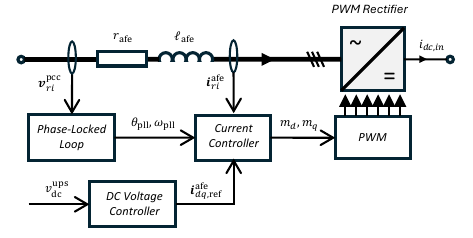}
     \caption{The control and circuit diagram of a three-phase PWM AFE rectifier}
     \label{fig:afe}
 \end{figure}

Fig.~\ref{fig:afe} illustrates the circuit and control architecture of a three-phase PWM AFE rectifier. The PCC voltage is processed by a phase-locked loop (PLL) to provide the synchronization angle and frequency for synchronous-frame control. An outer DC-link voltage PI loop regulates the DC-link voltage and generates the active current reference, while the reactive current reference is set to zero to enforce unity power factor. An inner $dq$-axis current PI loop with standard cross-coupling decoupling terms generates the converter voltage command, which is applied through PWM to shape the AC-side current and transfer power to the DC side. Neglecting switching ripple, the AFE DAEs are given in \eqref{eq:afe}.
 \begin{subequations}
\label{eq:afe}
\allowdisplaybreaks
\setlength{\jot}{2pt}
\thinmuskip=2mu
\medmuskip=2mu
\thickmuskip=2mu
\begin{align}
\shortintertext{\textit{Differential Equations}}
\frac{1}{\omega_b}\,\frac{\d \theta_\textnormal{pll}}{\d t}
&= \omega_\textnormal{pll}-\omega_s,
\\
\frac{\d \epsilon}{\d t}
&= v_q^\textnormal{pll},
\\
\frac{1}{\omega_\textnormal{lp}}\,\frac{\d v_q^\textnormal{pll}}{\d t}
&= \bm e_2^\top \bm v_{dq}^\textnormal{pcc}-v_q^\textnormal{pll},
\\
\frac{\ell_\textnormal{afe}}{\omega_b}\,\frac{\d \bm i_{dq}^\textnormal{afe}}{\d t}
&=
\bm v_{dq}^\textnormal{pcc}
-
v_\textnormal{dc}^\textnormal{ups}\,\bm m_{dq}
-
r_\textnormal{afe}\,\bm i_{dq}^\textnormal{afe}
\;+\;
\omega_\textnormal{pll}\,\ell_\textnormal{afe}\,\bm J\,\bm i_{dq}^\textnormal{afe},
\\
\frac{\d \xi_\textnormal{dc}^\textnormal{afe}}{\d t}
&= v_\textnormal{dc}^\textnormal{ups,ref}-v_\textnormal{dc}^\textnormal{ups},
\\
\frac{\d \bm \gamma^\textnormal{afe}_{dq}}{\d t}
&= \bm i_{dq}^\textnormal{afe}-\bm i_{dq,\textnormal{ref}}^\textnormal{afe}.
\\[2pt]
\shortintertext{\textit{Algebraic Equations}}
\omega_\textnormal{pll}
&= \omega_s + k_p^\textnormal{pll}\, v_q^\textnormal{pll} + k_i^\textnormal{pll}\, \epsilon,
\\
\bm v_{dq}^\textnormal{pcc}
&= \bm R(\theta_\textnormal{pll})\,\bm v_{ri}^\textnormal{pcc},
\\
\bm i_{ri}^\textnormal{pcc}
&= \bm R(\theta_\textnormal{pll})^\top \bm i_{dq}^\textnormal{afe},
\\
\bm i_{dq,\textnormal{ref}}^\textnormal{afe}
&=
\begin{bmatrix}
k_p^\textnormal{dc,afe}\!\left(v_\textnormal{dc}^\textnormal{ups,ref}-v_\textnormal{dc}^\textnormal{ups}\right)
+k_i^\textnormal{dc,afe}\,\xi_\textnormal{dc}^\textnormal{afe}
\\
0
\end{bmatrix},
\\
\bm v_{dq,\textnormal{ref}}^\textnormal{afe}
&=
k_p^\textnormal{c,afe}\!\left(\bm i_{dq}^\textnormal{afe}-\bm i_{dq,\textnormal{ref}}^\textnormal{afe}\right)
\!+\!
k_i^\textnormal{c,afe}\,\bm \gamma^\textnormal{afe}_{dq}
\!+\!
\omega_\textnormal{pll}\,\ell_\textnormal{afe}\,\bm J\,\bm i_{dq}^\textnormal{afe},
\\
\bm m_{dq}
&= \frac{1}{v_\textnormal{dc}^\textnormal{ups}}\,\bm v_{dq,\textnormal{ref}}^\textnormal{afe}.
\end{align}
\end{subequations}
\textit{Definition of Symbols}
\begin{itemize}
\item \textbf{Vector notation convention:}
$\bm x_{dq}\triangleq [x_d \,\, x_q]^\top$ and $\bm x_{ri}\triangleq [x_r \,\, x_i]^\top$ are quantities in the $dq$ frame established by the PLL and the stationary $ri$ frame defined in the grid reference, respectively.
\item \textbf{Operators and mappings:}
\[
\bm e_2 \triangleq \begin{bmatrix}0 & 1\end{bmatrix}^\top,  \,
\bm J \triangleq \begin{bmatrix}0&-1\\[1pt]1&0\end{bmatrix}, \,
\bm R(\theta)
\triangleq
\begin{bmatrix}
\cos\theta & \sin\theta\\
-\sin\theta & \cos\theta
\end{bmatrix}.
\]

\item \textbf{Parameters:}
$\omega_b$ is the base angular frequency and $\omega_s$ is the synchronous reference frequency; $\omega_\textnormal{lp}$ is the PLL low-pass filter bandwidth; $k_p^\textnormal{pll}$ and $k_i^\textnormal{pll}$ are the PLL gains; $\ell_\textnormal{afe}$ and $r_\textnormal{afe}$ are the AFE AC-side inductance and resistance; $v_\textnormal{dc}^\textnormal{ups,ref}$ is the DC-link voltage reference; $(k_p^\textnormal{dc,afe},k_i^\textnormal{dc,afe})$ are the DC-voltage-loop gains; and $(k_p^\textnormal{c,afe},k_i^\textnormal{c,afe})$ are the current-loop gains.
\item \textbf{Variables:}
$\theta_\textnormal{pll}$ is the PLL angle, $\omega_\textnormal{pll}$ is the PLL frequency estimate, $v_q^\textnormal{pll}$ is the filtered $q$-axis PCC voltage, and $\epsilon$ is the PLL integrator state; $\bm v_{ri}^\textnormal{pcc}$ and $\bm v_{dq}^\textnormal{pcc}$ are the PCC voltage vectors in the $ri$ and $dq$ frames, $\bm i_{ri}^\textnormal{pcc}$ is the PCC current vector in the $ri$ frame, and $\bm i_{dq}^\textnormal{afe}$ is the AFE AC-side current vector in the $dq$ frame; $v_\textnormal{dc}^\textnormal{ups}$ is the DC-link voltage; $\xi_\textnormal{dc}^\textnormal{afe}$ is the DC-voltage-loop integrator state; $\bm \gamma_{dq}^\textnormal{afe}$ is the current-loop integrator-state vector; $\bm i_{dq,\textnormal{ref}}^\textnormal{afe}$ is the current-reference vector; $\bm v_{dq,\textnormal{ref}}^\textnormal{afe}$ is the converter voltage-command vector; and $\bm m_{dq}$ is the modulation-index vector.
\end{itemize}

\subsection{Voltage Source Inverter}
\begin{figure}[!tb]
     \centering
     \includegraphics[width=1\linewidth]{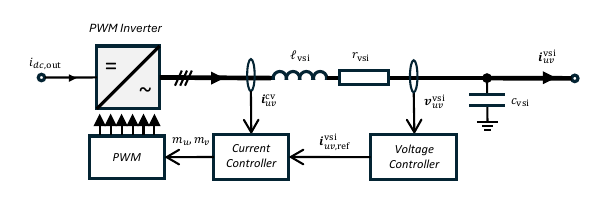}
     \caption{The control and circuit diagram of a three-phase PWM VSI}
     \label{fig:vsi}
 \end{figure}

Fig.~\ref{fig:vsi} shows the circuit and control architecture of the PWM VSI. The inverter converts the DC-link voltage to a regulated AC output and interfaces with the AC bus through an $RL$ filter branch and a shunt capacitor. The VSI adopts its own synchronous frame with fixed frequency $\omega_\textnormal{vsi}=\omega_s$, and regulates the capacitor voltage to a direct-axis-aligned reference (with zero quadrature-axis reference). An outer voltage PI loop generates the converter-side inductor current reference, and an inner current PI loop with standard rotational decoupling generates the converter voltage command (modulation command) for PWM. Neglecting switching ripple, the VSI DAEs are given in \eqref{eq:vsi}.
\begin{subequations}
\label{eq:vsi}
\allowdisplaybreaks
\setlength{\jot}{2pt}
\thinmuskip=2mu
\medmuskip=2mu
\thickmuskip=2mu
\begin{align}
\shortintertext{\textit{Differential Equations}}
\frac{\ell_\textnormal{vsi}}{\omega_b}\,\frac{\d \bm i_{uv}^\textnormal{cv}}{\d t}
&=
v_\textnormal{dc}^\textnormal{ups}\,\bm m_{uv}
-
\bm v_{uv}^\textnormal{vsi}
-
r_\textnormal{vsi}\,\bm i_{uv}^\textnormal{cv}
\;+\;
\omega_\textnormal{vsi}\ell_\textnormal{vsi}\,\bm J\,\bm i_{uv}^\textnormal{cv},
\\
\frac{c_\textnormal{vsi}}{\omega_b}\,\frac{\d \bm v_{uv}^\textnormal{vsi}}{\d t}
&=
\bm i_{uv}^\textnormal{cv}
-
\bm i_{uv}^\textnormal{vsi}
\;+\;
\omega_\textnormal{vsi}c_\textnormal{vsi}\,\bm J\,\bm v_{uv}^\textnormal{vsi},
\\
\frac{\d \bm \xi_{uv}^\textnormal{vsi}}{\d t}
&=
\bm v_{uv,\textnormal{ref}}^\textnormal{vsi}-\bm v_{uv}^\textnormal{vsi},
\\
\frac{\d \bm \gamma_{uv}^\textnormal{vsi}}{\d t}
&=
\bm i_{uv,\textnormal{ref}}^\textnormal{cv}-\bm i_{uv}^\textnormal{cv}.
\\
\shortintertext{\textit{Algebraic Equations}}
\omega_\textnormal{vsi}
&= \omega_s,
\\
\bm v_{uv,\textnormal{ref}}^\textnormal{vsi}
&= \begin{bmatrix} v_{u,\textnormal{ref}}^\textnormal{vsi}\\ 0 \end{bmatrix},
\\
\bm i_{uv,\textnormal{ref}}^\textnormal{cv}
&=
k_p^\textnormal{v,vsi}\!\big(\bm v_{uv,\textnormal{ref}}^\textnormal{vsi}-\bm v_{uv}^\textnormal{vsi}\big)
+k_i^\textnormal{v,vsi}\,\bm \xi_{uv}^\textnormal{vsi}
-\omega_\textnormal{vsi}c_\textnormal{vsi}\,\bm J\,\bm v_{uv}^\textnormal{vsi},
\\
\bm v_{uv,\textnormal{ref}}^\textnormal{cv}
&=
k_p^\textnormal{c,vsi}\!\big(\bm i_{uv,\textnormal{ref}}^\textnormal{cv}-\bm i_{uv}^\textnormal{cv}\big)
+k_i^\textnormal{c,vsi}\,\bm \gamma_{uv}^\textnormal{vsi}
-\omega_\textnormal{vsi}\ell_\textnormal{vsi}\,\bm J\,\bm i_{uv}^\textnormal{cv},
\\
\bm m_{uv}
&= \frac{1}{v_\textnormal{dc}^\textnormal{ups}}\,\bm v_{uv,\textnormal{ref}}^\textnormal{cv}.
\end{align}
\end{subequations}
\textit{Definition of Symbols}
\begin{itemize}
\item \textbf{Vector notation convention:}
$\bm x_{uv}\triangleq [x_u \,\, x_v]^\top$ denotes an quantity in the $uv$ frame established by the VSI.

\item \textbf{Parameters:}
$\ell_\textnormal{vsi}$ and $r_\textnormal{vsi}$ are the VSI filter inductance and resistance; $c_\textnormal{vsi}$ is the shunt capacitance; $v_{u,\textnormal{ref}}^\textnormal{vsi}$ is the $u$-axis voltage reference with $v_{v,\textnormal{ref}}^\textnormal{vsi}\equiv 0$; $(k_p^\textnormal{v,vsi},k_i^\textnormal{v,vsi})$ are the outer voltage-loop gains; and $(k_p^\textnormal{c,vsi},k_i^\textnormal{c,vsi})$ are the inner current-loop gains.

\item \textbf{Variables:}
$\omega_\textnormal{vsi}$ is the VSI synchronous speed (set to $\omega_s$ here); $\bm i_{uv}^\textnormal{cv}$ is the converter-side (inductor) current vector, $\bm v_{uv}^\textnormal{vsi}$ is the filter-capacitor voltage vector, and $\bm i_{uv}^\textnormal{vsi}$ is the VSI output current vector; $\bm \xi_{uv}^\textnormal{vsi}$ and $\bm \gamma_{uv}^\textnormal{vsi}$ are the integrator-state vectors of the outer voltage loop and inner current loop; $\bm v_{uv,\textnormal{ref}}^\textnormal{vsi}$ is the voltage-reference vector, $\bm i_{uv,\textnormal{ref}}^\textnormal{cv}$ is the current-reference vector, $\bm v_{uv,\textnormal{ref}}^\textnormal{cv}$ is the converter voltage-command vector, and $\bm m_{uv}$ is the modulation-index vector.
\end{itemize}

\subsection{DC-Link Capacitor}
The AFE rectifier and VSI are coupled through a shared DC bus with a capacitor. The DC-link voltage dynamics are governed by the current imbalance between the DC current injected by the AFE and the DC current drawn by the VSI. Neglecting switching ripple, the DC-link DAE model is given in \eqref{eq:dclink}.
\begin{subequations}
\label{eq:dclink}
\allowdisplaybreaks
\setlength{\jot}{2pt}
\thinmuskip=2mu
\medmuskip=2mu
\thickmuskip=2mu
\begin{align}
\shortintertext{\textit{Differential Equation}}
\frac{c_\textnormal{dc}}{\omega_b}\,\frac{\d v_\textnormal{dc}^\textnormal{ups}}{\d t}
&=
i_\textnormal{dc,in}-i_\textnormal{dc,out}.
\\[2pt]
\shortintertext{\textit{Algebraic Equations}}
i_\textnormal{dc,in}
&= \bm m_{dq}^\top \bm i_{dq}^\textnormal{afe},
\\
i_\textnormal{dc,out}
&= \bm m_{uv}^\top \bm i_{uv}^\textnormal{cv}.
\end{align}
\end{subequations}
\textit{Definition of Symbols}
\begin{itemize}
\item \textbf{Parameters:}
$c_\textnormal{dc}$ is the DC-link capacitance.

\item \textbf{Variables:}
$i_\textnormal{dc,in}$ and $i_\textnormal{dc,out}$ are the DC-link capacitor input and output currents, respectively.
\end{itemize}

\subsection{Array of Power Supply Units}
\begin{figure}[!tb]
     \centering
     \includegraphics[width=0.85\linewidth]{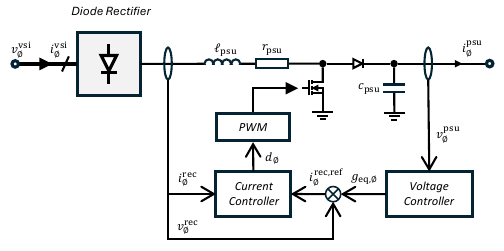}
     \caption{The control and circuit diagram of a single-phase PSU}
     \label{fig:psu}
 \end{figure}

The PSU-array model is derived by modeling a single-phase PSU and stacking three identical phase models. Fig.~\ref{fig:psu} shows a typical \textbf{single-phase} PSU, modeled as a rectifier--boost PFC stage connecting the upstream AC bus and the downstream DC port. The PSU is assumed to operate in continuous-conduction mode. It employs a cascaded control structure: an outer voltage PI loop regulates the PSU DC-port voltage and generates an equivalent conductance command, which is multiplied by the rectified input-voltage  to form the input-current reference (thus enforcing PFC-like resistive input behavior); an inner current PI loop tracks this current reference and outputs the PWM duty-ratio command.

While stacking three phase models provides a transparent component-level description, it is not convenient for oscillation analysis because the composite system is represented in mixed coordinates: the VSI is naturally modeled in the synchronous $uv$ frame, whereas the PSU array is in the phase domain. Coupling the two through Clarke/Park transforms introduces explicit time dependence via the VSI angle $\theta_\textnormal{vsi}=\omega_b\omega_\textnormal{vsi}t$, which makes the composite model time-periodic even in steady operation. In addition, single-phase PFC conversion inherently produces a $2\omega$ ripple on each PSU DC-side capacitor voltage, causing small phase-to-phase mismatches.

To obtain a time-invariant dynamic equivalent suitable for oscillation analysis, we adopt the following assumptions:
 \begin{itemize}
  \item \emph{Balanced three-phase operation:} data-center distribution is typically phase-interleaved and capacity-balanced, so the aggregated PSU array seen from the VSI AC bus is predominantly positive-sequence; residual unbalance is treated as a small perturbation and neglected.
  \item \emph{Negligible DC-side $2\omega$ ripple:} while a single-phase PFC stage introduces $2\omega$ power pulsation, the ripple is buffered by DC-side capacitance and is further smoothed by the aggregation of many downstream converters and loads, making it small relative to the nominal DC voltage.
  \item \emph{Quasi-steady-state inner-current loop:} the inner current loop of  boost-PFC is typically in the kHz range, while the outer voltage loop is much slower (typically a few--tens of Hz and often well  below $2\omega$), which justifies a quasi-steady-state (QSS) approximation for model-order reduction without materially affecting low-frequency modes.
\end{itemize}
Under these assumptions, the three-phase PSU-array DC ports are aggregated into a common DC-side model, and the AC-side dynamics are expressed directly in the VSI-synchronous $uv$ frame, yielding the simplified DAEs in \eqref{eq:psu_uv_qss_xi_state}. The detailed derivation is provided in \ref{app:psu} of the Supplementary Material.
\begin{subequations}
\label{eq:psu_uv_qss_xi_state}
\allowdisplaybreaks
\setlength{\jot}{2pt}
\thinmuskip=2mu
\medmuskip=2mu
\thickmuskip=2mu
\begin{align}
\shortintertext{\textit{Differential Equations}}
\frac{c_\textnormal{psu}}{\omega_b}\,
\frac{\d v^{\textnormal{psu}}}{\d t}
&=
\frac{
\Big(g_\textnormal{eq}-r_\textnormal{psu}g_\textnormal{eq}^2\Big)\,
\bigl\lVert \bm v_{uv}^\textnormal{vsi}\bigr\rVert_2^2
}{3v^{\textnormal{psu}}}
-
\,i^{\textnormal{psu}},
\\
\frac{\d \xi^{\textnormal{psu}}}{\d t}
&=
v^{\textnormal{psu,ref}}-v^{\textnormal{psu}}.
\\[2pt]
\shortintertext{\textit{Algebraic Equations}}
g_\textnormal{eq}
&=
k_p^\textnormal{v,psu}\,
\bigl(v^{\textnormal{psu,ref}}-v^{\textnormal{psu}}\bigr)
+
k_i^\textnormal{v,psu}\,\xi^{\textnormal{psu}},
\\
\bm i_{uv}^\textnormal{vsi}
&=
g_\textnormal{eq}\,\bm v_{uv}^\textnormal{vsi}.
\end{align}
\end{subequations}
\textit{Definition of Symbols}
\begin{itemize}
\item \textbf{Operators and mappings:}
$\|\bm x\|_2$ denotes vector $l_2$ norm

\item \textbf{Parameters:}
$c_\textnormal{psu}$ is the (per-phase) PSU DC-port capacitance; $r_\textnormal{psu}$ is the (per-phase) equivalent series resistance capturing the PSU conduction loss; $(k_p^\textnormal{v,psu},k_i^\textnormal{v,psu})$ are the gains of the PSU outer voltage PI controller; and $v^{\textnormal{psu,ref}}$ is the PSU DC-port voltage reference.

\item \textbf{Variables:}
$v^{\textnormal{psu}}$ is the aggregated PSU DC-port voltage (i.e., common DC component); $\xi^{\textnormal{psu}}$ is the integrator state of the PSU outer voltage loop; $g_\textnormal{eq}$ is the scalar equivalent input conductance commanded by the PSU voltage loop; and $i^{\textnormal{psu}}$ is the aggregated DC current drawn from the PSU DC port to the downstream DC--DC converter and load equivalent.
\end{itemize}

 \subsection{Downstream DC--DC Converter and Load Equivalent}
 \begin{figure}[!tb]
     \centering
     \includegraphics[width=0.6\linewidth]{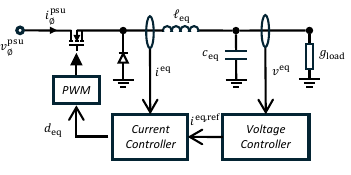}
     \caption{The control and circuit diagram of per-phase downstream DC--DC converter and load equivalent }
     \label{fig:dcdc_eq}
 \end{figure}
Fig.~\ref{fig:dcdc_eq} shows a per-phase equivalent of the downstream data-center power stage, which aggregates the fast-switching DC--DC conversion chain (e.g., isolated DC--DC converters and voltage-regulation modules) and the IT load into a single buck-converter equivalent with cascaded voltage--current control. The outer voltage loop regulates the load-side DC voltage to a prescribed reference, while the inner current loop shapes the converter input current drawn from the upstream PSU DC link. The aggregated load conductance is parameterized by the total load power at the regulated voltage level, so that the equivalent captures the dominant low-frequency power/voltage dynamics seen by the upstream AC/DC interface without resolving the switching details of numerous parallel converters and loads.

As in the PSU model, a clear time-scale separation between the outer voltage loop and the faster inner current loop motivates a QSS approximation for the inner loop. The resulting reduced-order DAEs are given in \eqref{eq:dcdc_qss_xi_state}. The detailed derivation is provided in \ref{app:eq} of the Supplementary Material.
\begin{subequations}
\label{eq:dcdc_qss_xi_state}
\allowdisplaybreaks
\setlength{\jot}{2pt}
\thinmuskip=2mu
\medmuskip=2mu
\thickmuskip=2mu
\begin{align}
\shortintertext{\textit{Differential Equations}}
\frac{c_\textnormal{eq}}{\omega_b}\,
\frac{\d v^\textnormal{eq}}{\d t}
&=
\,i^\textnormal{eq}
-
\,g_\textnormal{load}\,v^\textnormal{eq},
\\[2pt]
\frac{\d \xi^\textnormal{eq}}{\d t}
&=
v^{\textnormal{eq,ref}}-v^\textnormal{eq}.
\\[2pt]
\shortintertext{\textit{Algebraic Equations}}
g_\textnormal{load}
&=
\frac{p_\textnormal{load}}{3\left(v^{\textnormal{eq,ref}}\right)^2},
\\[2pt]
i^{\textnormal{eq}}
&=
k_p^\textnormal{v,eq}
\bigl(
v_o^{\textnormal{eq,ref}}
-
v_o^\textnormal{eq}
\bigr)
+
k_i^\textnormal{v,eq}\,
\xi^\textnormal{eq},\\ 
i^\textnormal{psu}
&=
\frac{v^\textnormal{eq}}{v^\textnormal{psu}}\,
i^\textnormal{eq}.
\end{align}
\end{subequations}
\textit{Definition of Symbols}
\begin{itemize}
\item \textbf{Parameters:}
$c_\textnormal{eq}$ is the equivalent output capacitance of the aggregated downstream DC--DC/load stage; $v^{\textnormal{eq,ref}}$ is the regulated load-side voltage reference; $p_\textnormal{load}$ is the total three-phase load power; and $g_\textnormal{load}$ is the per-phase equivalent load conductance defined by $p_\textnormal{load}$ at the regulated voltage level.

\item \textbf{Variables:}
$v^\textnormal{eq}$ is the aggregated load-side (regulated) DC voltage; $\xi^\textnormal{eq}$ is the integrator state of the outer voltage loop; $i^\textnormal{eq}$ is the commanded load-side current generated by the voltage loop (under the inner current-loop QSS approximation).
\end{itemize}

\section{Power-System Oscillation Analysis}
\label{sec:osc}
Based on the component-level model derived in Section~\ref{sec:dae}, the complete data-center power-delivery chain can be represented as a nonlinear DAE system with input. When jointly formulated with the grid-side DAEs (e.g., transmission lines and generators), the overall interconnected system admits a closed, time-invariant representation. This section develops a small-signal oscillation analysis framework for the composite model under the online operating mode, with a focus on how control-loop interactions across the AFE, DC link, VSI, PSU array, downstream DC--DC converter, and load stages produce lightly damped modes. The framework also clarifies how load-side power variations propagate upstream into the grid side and excite these modes, leading to oscillation amplification.

\subsection{Composite DAE Model}
The DAE model of the complete grid-connected data-center power-delivery chain is represented in the composite form
\begin{subequations}
\label{eq:DAE}
\begin{align}
    \frac{\d \bm x }{\d t} &= \bm f(\bm x, \bm y, w ; \bm p), \\
    \bm 0 &= \bm g(\bm x, \bm y, w ; \bm p).
\end{align}
\end{subequations}
where $\bm x$ contains all differential states appearing in \eqref{eq:afe}--\eqref{eq:dcdc_qss_xi_state}, together with the differential states of the grid-side dynamic components, and $\bm y$ collects the corresponding algebraic variables. The scalar $w$ denotes an exogenous input/disturbance (e.g., load-power variation), and $\bm p$ collects the model parameters.

\subsection{Small-Signal Modeling}
When the overall system operates at a steady state, denoted by $(\bm x_0,\bm y_0)$ under a given exogenous input/disturbance $w_0$ and parameter set $\bm p$, we enforce
\[
\frac{\d \bm x}{\d t}=\bm 0.
\]
Substituting $(\bm x_0,\bm y_0,w_0)$ into \eqref{eq:DAE} yields the equilibrium equations
\begin{subequations}
\label{eq:DAE_ss}
\begin{align}
\bm 0 &= \bm f(\bm x_0,\bm y_0,w_0;\bm p),\\
\bm 0 &= \bm g(\bm x_0,\bm y_0,w_0;\bm p).
\end{align}
\end{subequations}
The solution of \eqref{eq:DAE_ss} defines the operating point for the subsequent small-signal oscillation analysis. Define small perturbations around the operating point as $\Delta \bm x \triangleq \bm x-\bm x_0$, $\Delta \bm y \triangleq \bm y-\bm y_0$, and $\Delta w \triangleq w-w_0$. Linearizing \eqref{eq:DAE} around $(\bm x_0,\bm y_0,w_0)$ gives the following small-signal DAE model
\begin{subequations}
\label{eq:DAE_lin}
\begin{align}
 \frac{\d {\Delta\bm x}}{ \d t}
&=
\bm f_x\,\Delta \bm x
+\bm f_y\,\Delta \bm y
+\bm f_w\,\Delta w,
\\
\bm 0
&=
\bm g_x\,\Delta \bm x
+\bm g_y\,\Delta \bm y
+\bm g_w\,\Delta w,
\end{align}
\end{subequations}
where the Jacobians are evaluated at the operating point, i.e., $\bm f_x \triangleq \left.\partial \bm f/\partial \bm x\right|_0$, $\bm f_y \triangleq \left.\partial \bm f/\partial \bm y\right|_0$, $\bm f_w \triangleq \left.\partial \bm f/\partial w\right|_0$, $\bm g_x \triangleq \left.\partial \bm g/\partial \bm x\right|_0$, $\bm g_y \triangleq \left.\partial \bm g/\partial \bm y\right|_0$, and $\bm g_w \triangleq \left.\partial \bm g/\partial w\right|_0$. Assuming $\bm g_y$ is nonsingular in a neighborhood of the operating point, the algebraic perturbation can be eliminated as
\begin{align*}
\Delta \bm y
=
-\bm g_y^{-1}\bm g_x\,\Delta \bm x
-\bm g_y^{-1}\bm g_w\,\Delta w.
\end{align*}
Substituting $\Delta \bm y$ into \eqref{eq:DAE_lin} yields the reduced small-signal state-space model
\begin{align}
\label{eq:ss_lin}
\frac{\d {\Delta\bm x}}{ \d t}
&=
\bm A\,\Delta \bm x+\bm b\,\Delta w.
\end{align}
where $\bm A \triangleq \bm f_x-\bm f_y\bm g_y^{-1}\bm g_x$, and $\bm b \triangleq \bm f_w-\bm f_y\bm g_y^{-1}\bm g_w$.

\subsection{Modal Analysis and Participation Factors}
Based on the reduced small-signal model \eqref{eq:ss_lin}, the oscillatory behavior of the interconnected system is characterized by the eigenstructure of the state matrix $\bm A$. Let $\lambda_k$ denote the $k^\textnormal{th}$ eigenvalue of $\bm A$, with the associated left and right eigenvectors $\bm l_k$ and $\bm r_k$ satisfying
\begin{align}
\bm l_k^\top \bm A = \lambda_k \bm l_k^\top, \qquad
\bm A \bm r_k = \lambda_k \bm r_k.
\end{align}
The left and right eigenvectors are normalized such that
\begin{align}
\bm l_k^\top \bm r_k = 1.
\end{align}
To quantify the contribution of each state to a given mode, the participation factor of the $i^\textnormal{th}$ state in the $k^\textnormal{th}$ mode is defined as
\begin{align}
\rho_{ik} \triangleq \Re\!\left\{l_{k,i} r_{i,k}\right\},
\label{eq:pf}
\end{align}
where $l_{k,i}$ and $r_{i,k}$ are the $i^\textnormal{th}$ entries of $\bm l_k$ and $\bm r_k$, respectively, and $\Re\{\cdot\}$ denotes the real-part operator. The participation factors in \eqref{eq:pf} are used to rank the dominant state variables of each critical mode (ordered by $|\rho_{ik}|$), thereby providing an interpretable way to trace each weakly damped mode to its root cause across the data-center power-delivery chain and the grid-side network.

\subsection{Power Oscillation Amplification Analysis}
To quantify how server-load  fluctuations induced by time-varying computing demand propagate through the data-center power-delivery chain, we introduce a new metric named  power oscillation amplification (POA) factor, which is defined as the gain of the transfer function from the server-load disturbance $\Delta p_\textnormal{load}\triangleq p_\textnormal{load}-p_{\textnormal{load},0}$ to the PCC active-power ripple $\Delta p_\textnormal{pcc}\triangleq p_\textnormal{pcc}-p_{\textnormal{pcc},0}$. We choose $p_\textnormal{pcc}$ as the output variable:
\begin{align}
z \triangleq p_\textnormal{pcc}
= h(\bm x,\bm y)
= \left(\bm v_{dq}^\textnormal{pcc}\right)^\top \bm i_{dq}^\textnormal{pcc}.
\label{eq:pcc_power_output}
\end{align}
Linearizing \eqref{eq:pcc_power_output} around the operating point gives
\begin{align}
\Delta z
=
\bm h_x\,\Delta \bm x+\bm h_y\,\Delta \bm y,
\label{eq:dz_lin}
\end{align}
where $\bm h_x \triangleq \left.\partial h/\partial \bm x\right|_0$ and $\bm h_y \triangleq \left.\partial h/\partial \bm y\right|_0$.
Similar to \eqref{eq:ss_lin}, eliminating the algebraic perturbation $\Delta \bm y$ yields
\begin{align}
\Delta p_\textnormal{pcc}
=
\Delta z
=
\bm c\,\Delta \bm x,
\label{eq:pcc_out_red}
\end{align}
where $\bm c \triangleq \bm h_x-\bm h_y\bm g_y^{-1}\bm g_x$. Based on \eqref{eq:ss_lin} and \eqref{eq:pcc_out_red}, the small-signal transfer function from the load disturbance $\Delta p_\textnormal{load}$ (i.e., $\Delta w$) to the PCC active-power response is:
\begin{align}
G_{\textnormal{pcc}\leftarrow \textnormal{load}}(s)
\triangleq
\frac{\Delta p_\textnormal{pcc}(s)}{\Delta p_\textnormal{load}(s)}
=
\bm c\,(s\bm I-\bm A)^{-1}\bm b.
\label{eq:G_pcc_load}
\end{align}
Accordingly, the POA factor is defined as
\begin{align}
\mathrm{POA}(\omega)
\triangleq
\left|G_{\textnormal{pcc}\leftarrow \textnormal{load}}(j\omega)\right|.
\label{eq:POA_def}
\end{align}
Here, a large \(\mathrm{POA}(\omega)\)  at frequency \(\omega=\omega_\textnormal{peak}\) indicates a strong amplification of load-side power oscillations as they propagate to the grid. Since \(\mathrm{POA}(\omega)=|G_{\textnormal{pcc}\leftarrow \textnormal{load}}(j\omega)|\) depends on both \(\bm A\) and the coupling terms \((\bm b,\bm c)\), it reflects both intrinsic modal properties and input--output coupling. To further quantify the contribution of each mode to the POA, we introduce the modal residue of the \(k^\textnormal{th}\) mode as
\begin{align}
\bm R_k \triangleq (\bm c\bm r_k)(\bm l_k^\top \bm b),
\end{align}
Using the partial-fraction expansion of \((j\omega\bm I-\bm A)^{-1}\), the POA can be decomposed into modal components as
\begin{align}
\mathrm{POA}(\omega)
=
\left|
\sum_{k=1}^{n}
\frac{\bm R_k}{j\omega-\lambda_k}
\right|.
\label{eq:POA_modal_residue}
\end{align}
Thereby, \(|\bm R_k|\) is used to indicate the overall excitability and observability of each mode in the POA channel. Note that some modes with eigenvalues close to the imaginary axis may neither be excited by server-load fluctuations nor observed from the grid variables. Such modes therefore contribute little to the observed oscillation phenomena.

\section{Numerical Results}
\label{sec:case}
In this section, two case studies are presented. The first case considers a single data-center system connected to an infinite bus and is used to validate the proposed model, investigate the intrinsic power-oscillation amplification mechanisms within the data-center power-delivery chain, and simulate the PCC power profile driven by real-world GPU-cluster workload. The second case connects the data-center model to a modified 3-machine 9-bus system, in which the three generation units are modeled by a synchronous generator, a grid-forming converter, and a grid-following converter, respectively. This case is used to examine how load-side power variations interact with grid-side electromechanical transients, together with the coupling among heterogeneous power-electronic devices.
\subsection{Single Data Center Infinite Bus (SDCIB) System}
\begin{table}[!b]
\centering
\caption{Default Parameter Setup for the SDCIB Case}
\label{tab:sdcib_params}
\renewcommand{\arraystretch}{1}
\setlength{\tabcolsep}{0pt}
\begin{tabular}{llp{5cm}}
\toprule
\textbf{Symbol} & \textbf{Value} & \textbf{Description} \\
\midrule
$\omega_b$ & $2\pi\times 60~\textnormal{rad/s}$ & Base angular frequency \\
$\omega_s$ & $1.0~\textnormal{p.u.}$ & Grid synchronous speed \\
$\omega_\textnormal{lp}$ & $2\pi\times 100~\textnormal{rad/s}$ & Low-pass filter cutoff frequency \\

$V_\infty$ & $1.0~\textnormal{p.u.}$ & Infinite-bus voltage magnitude \\
$R_\infty$, $X_\infty$ & $0.02$,    $0.19~\textnormal{p.u.}$ & Grid impedance \\

$r_\textnormal{afe}$, $\ell_\textnormal{afe}$ & $0.003$, $0.05~\textnormal{p.u.}$ & AFE filter impedance  \\
$c_\textnormal{dc}$ & $2.0~\textnormal{p.u.}$ & UPS DC-link capacitance \\

$k_p^\textnormal{pll}$, $k_i^\textnormal{pll}$ & $0.471$, $41.89$ & PLL ($f_{\mathrm{bw}}=20~\textnormal{Hz},\,\zeta=0.707$) \\

$k_p^\textnormal{dc,afe}$, $k_i^\textnormal{dc,afe}$  & $0.333$, $5.236$  & AFE DC-PI ($f_{\mathrm{bw}}=5~\textnormal{Hz},\,\zeta=1.0$) \\

$k_p^\textnormal{c,afe}$, $k_i^\textnormal{c,afe}$ & $0.233$, $209.4$ & AFE current-PI ($f_{\mathrm{bw}}=200~\textnormal{Hz},\zeta=0.707$) \\

$r_\textnormal{vsi}$, $\ell_\textnormal{vsi}$ & $0.003$, $0.05~\textnormal{p.u.}$ & VSI filter impedance \\
$c_\textnormal{vsi}$ & $0.2~\textnormal{p.u.}$ & VSI filter capacitance \\

$v_\textnormal{dc}^\textnormal{ups,ref}$ & $1.0~\textnormal{p.u.}$ & UPS DC-link voltage reference \\
$v_{u,\textnormal{ref}}^\textnormal{vsi}$ & $1.0~\textnormal{p.u.}$ & VSI AC-voltage reference \\

$k_p^\textnormal{v,vsi}$, $k_i^\textnormal{v,vsi}$ & $0.667$, $209.4$ & VSI voltage-PI ($f_{\mathrm{bw}}=100~\textnormal{Hz},\,\zeta=1.0$) \\

$k_p^\textnormal{c,vsi}$, $k_i^\textnormal{c,vsi}$ & $0.664$, $837.8$ & VSI current-PI ($f_{\mathrm{bw}}=400~\textnormal{Hz},\,\zeta=1.0$) \\

$c_\textnormal{psu}$ & $2.0~\textnormal{p.u.}$ & PSU DC-port capacitance \\
$r_\textnormal{psu}$ & $0.005~\textnormal{p.u.}$ & PSU equivalent resistance \\
$v^{\textnormal{psu,ref}}$ & $1.0~\textnormal{p.u.}$ & PSU DC-voltage reference \\

$k_p^\textnormal{v,psu}$, $k_i^\textnormal{v,psu}$ & $0.667$, $20.94$ & PSU voltage-PI ($f_{\mathrm{bw}}=10~\textnormal{Hz},\,\zeta=1.0$) \\

$c_\textnormal{eq}$ & $0.2~\textnormal{p.u.}$ & Equivalent load-side capacitance \\
$v^{\textnormal{eq,ref}}$ & $0.5~\textnormal{p.u.}$ & Load-side voltage reference \\

$k_p^\textnormal{v,eq}$, $k_i^\textnormal{v,eq}$ & $0.667$, $209.4$ & Load-side DC-PI ($f_{\mathrm{bw}}=100~\textnormal{Hz},\,\zeta=1.0$) \\

$p_\textnormal{load}$ & $0.5~\textnormal{p.u.}$ & Server load demand \\

\bottomrule
\end{tabular}
\end{table}

\subsubsection{Parameter Setting} 
In this case, the grid is represented by an ideal infinite-bus voltage source behind an equivalent series impedance, and the data-center power-supply model derived in Section~\ref{sec:dae} is connected at the PCC. The network-side variables are expressed in the grid's stationary $ri$ frame, and the infinite-bus voltage is aligned with the $r$ axis, i.e., $\bm v_{ri}^{\infty} = [V_\infty \,\, 0]^\top$, where $V_\infty$ is the infinite-bus voltage magnitude. The PCC voltage is given by
\begin{align}
\bm v_{ri}^{\textnormal{pcc}}
=
\bm v_{ri}^{\infty}
-
\begin{bmatrix}
R_\infty & -X_\infty\\
X_\infty & \ \ R_\infty
\end{bmatrix}\,\bm i_{ri}^{\textnormal{afe}},
\label{eq:pcc_infbus}
\end{align}
The algebraic equation \eqref{eq:pcc_infbus} provides the grid-side closure for the proposed model \eqref{eq:afe}--\eqref{eq:dcdc_qss_xi_state} in the SDCIB case and couples the infinite-bus network to the AFE through $\bm v_{ri}^{\textnormal{pcc}}$ and $\bm i_{ri}^{\textnormal{afe}}$.The default SDCIB parameters are listed in Table~\ref{tab:sdcib_params}. PI controller gains are  are obtained from a bandwidth-based tuning rule with target bandwidth $f_{\mathrm{bw}}$ and damping ratio $\zeta$. We refer \ref{app:tuning} of the Supplementary Material for details.

\begin{figure*}[!tb]
\centering

\begin{minipage}[t]{0.32\textwidth}
    \centering
    \includegraphics[width=\linewidth]{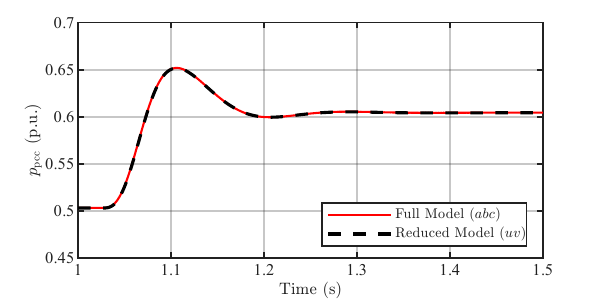}\\[-1mm]
    {\footnotesize (a)}
\end{minipage}\hfill
\begin{minipage}[t]{0.32\textwidth}
    \centering
    \includegraphics[width=\linewidth]{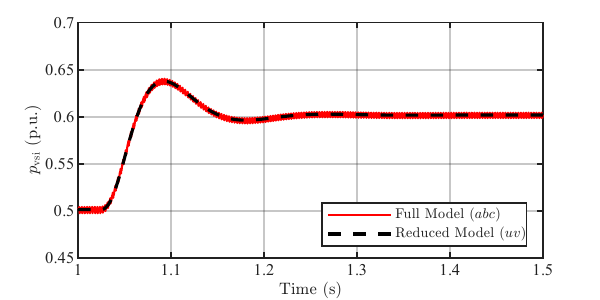}\\[-1mm]
    {\footnotesize (b)}
\end{minipage}\hfill
\begin{minipage}[t]{0.32\textwidth}
    \centering
    \includegraphics[width=\linewidth]{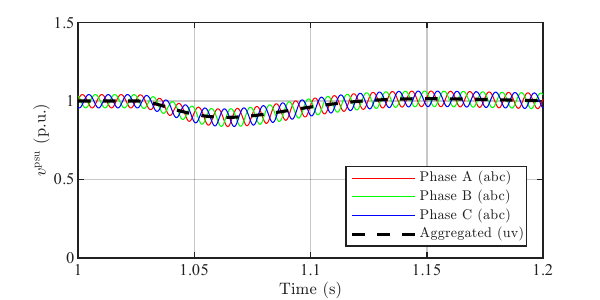}\\[-1mm]
    {\footnotesize (c)}
\end{minipage}

\caption{Comparison between the full-order three-phase model and the reduced model under a step change in $p_{\textnormal{load}}$ from 0.5 p.u. to 0.6 p.u.: (a) PCC active power, (b) VSI AC-bus active power, and (c) PSU DC-side voltage.}
\label{fig:qss_validation_step}
\end{figure*}

\begin{figure*}[!tb]
     \centering
    \includegraphics[width=\linewidth]{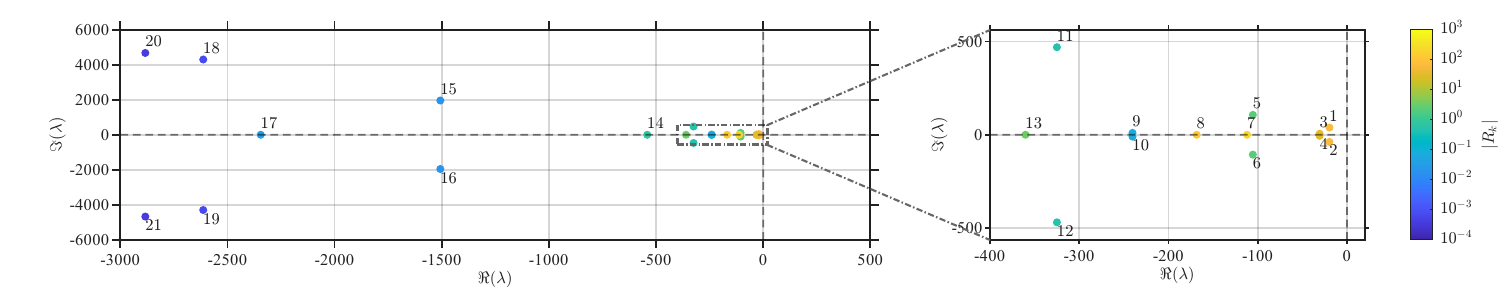}
     \caption{Eigenvalue map and mode residues of the SDCIB case}
     \label{fig:eigen}
 \end{figure*} 

\subsubsection{Model Validation} To verify the QSS and common DC-side approximations used to derive \eqref{eq:psu_uv_qss_xi_state} and \eqref{eq:dcdc_qss_xi_state} from the three-phase full-order models \eqref{eq:pfc_abc_dcside} and \eqref{eq:dcdc_buck_equiv_scalar_currloop} (see the Supplementary Material), the proposed model is compared with the full-order model using time-domain simulations of a step change in $p_\textnormal{load}$ from 0.5 p.u. to 0.6 p.u. Both models follows parameter setup in Table~\ref{tab:sdcib_params}, and the full-order model has extra parameters (see Table. \ref{tab:extra_params} in the Supplementary Material) regarding the reduced current-loop controllers. 
The comparative  results in the SDCIB case are shown in Fig.\ref{fig:qss_validation_step}. The $2\omega$ DC-side voltage ripple of PSU arrays is shown in Fig.~\ref{fig:qss_validation_step}(c), indicating that the common-DC approximation accurately captures the aggregated dynamic behavior of the three phases. This $2\omega$ ripple is not observed in the VSI AC-bus power $p_\textnormal{vsi} $ in Fig.~\ref{fig:qss_validation_step}(b), because it cancels out when summed over the three phases. A small 6$\omega$ ripple is shown in Fig.~\ref{fig:qss_validation_step}(b), which is caused by the tracking lag of the current controller around the zero-crossing points. Since this ripple is propagated and filtered through the DC-link capacitor of UPS, the PCC power responses of the two models are nearly identical as shown in Fig.~\ref{fig:qss_validation_step}(a).

\begin{table}[!b]
\centering
\caption{Participation Factors of All Modes in the SDCIB Case}
\label{tab:pf_all_modes_top3_compact_1col}
\renewcommand{\arraystretch}{1}
\setlength{\tabcolsep}{1pt}
\footnotesize
\begin{tabularx}{\linewidth}{c c c >{\raggedright\arraybackslash}X >{\raggedright\arraybackslash}X}
\toprule
\textbf{Mode(s)} & \textbf{$\lambda=\sigma\pm j\omega$} & \textbf{$f = \omega/(2\pi)$} &
\textbf{Top states} & \textbf{Participation} \\
\midrule
1/2  & $-19.7 \pm j\,38.6$  & 6.15 Hz &
$v^{\textnormal{psu}},\,\xi^{\textnormal{psu}},\,\xi_{u}^{\textnormal{vsi}}$ &
0.462,\,0.360,\,0.174 \\
\addlinespace[1pt]

3/4  & $-30.6 \pm j\,5.36$  & 0.852 Hz &
$v_{\textnormal{dc}}^{\textnormal{ups}},\,\xi_{\textnormal{dc}}^{\textnormal{afe}},\,\gamma_{d}^{\textnormal{afe}}$ &
0.500,\,0.496,\,0.003 \\
\addlinespace[1pt]

5/6  & $-105 \pm j\,107$    & 16.97 Hz &
$\theta_{\textnormal{pll}},\,\epsilon_{\textnormal{pll}},\,v_{q}^{\textnormal{pll}}$ &
0.452,\,0.372,\,0.127 \\
\addlinespace[1pt]

7     & $-112$               & -- &
$\xi_{u}^{\textnormal{vsi}},\,v^{\textnormal{psu}},\,\xi^{\textnormal{psu}}$ &
0.566,\,0.238,\,0.093 \\
\addlinespace[1pt]

8     & $-168$               & -- &
$\xi^{\textnormal{eq}},\,v^{\textnormal{eq}}$ &
0.933,\,0.067 \\
\addlinespace[1pt]

9/10 & $-240 \pm j\,9.99$   & 1.59 Hz &
$\xi_{v}^{\textnormal{vsi}},\,\gamma_{v}^{\textnormal{vsi}},\,v_{v}^{\textnormal{vsi}}$ &
0.496,\,0.385,\,0.115 \\
\addlinespace[1pt]

11/12 & $-325 \pm j\,470$   & 74.78 Hz &
$\gamma_{d}^{\textnormal{afe}},\,\gamma_{q}^{\textnormal{afe}},\,v_{q}^{\textnormal{pll}}$ &
0.315,\,0.311,\,0.097 \\
\addlinespace[1pt]

13    & $-360$               & -- &
$\gamma_{u}^{\textnormal{vsi}},\,\xi_{u}^{\textnormal{vsi}},\,v^{\textnormal{psu}}$ &
0.530,\,0.239,\,0.088 \\
\addlinespace[1pt]

14    & $-541$               & -- &
$v_{q}^{\textnormal{pll}},\,\gamma_{d}^{\textnormal{afe}},\,\theta_{\textnormal{pll}}$ &
0.640,\,0.115,\,0.103 \\
\addlinespace[1pt]

15/16 & $-1507 \pm j\,1956$ & 311 Hz &
$i_{d}^{\textnormal{afe}},\,i_{q}^{\textnormal{afe}},\,\gamma_{q}^{\textnormal{afe}}$ &
0.393,\,0.389,\,0.101 \\
\addlinespace[1pt]

17    & $-2345$              & -- &
$v^{\textnormal{eq}},\,\xi^{\textnormal{eq}}$ &
0.933,\,0.067 \\
\addlinespace[1pt]

18/19 & $-2613 \pm j\,4297$ & 684 Hz &
$i_{u}^{\textnormal{cv}},\,i_{v}^{\textnormal{cv}},\,v_{v}^{\textnormal{vsi}}$ &
0.255,\,0.254,\,0.188 \\
\addlinespace[1pt]

20/21 & $-2883 \pm j\,4672$ & 744 Hz &
$i_{u}^{\textnormal{cv}},\,i_{v}^{\textnormal{cv}},\,v_{v}^{\textnormal{vsi}}$ &
0.242,\,0.242,\,0.205 \\
\bottomrule
\end{tabularx}
\end{table}

\subsubsection{Load-Driven Oscillation Analysis}

\begin{figure*}[!tb]
\centering

\begin{minipage}[t]{0.33\textwidth}
    \centering
    \includegraphics[width=\linewidth]{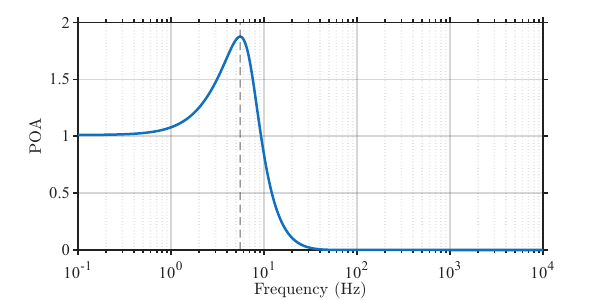}\\[-1mm]
    {\footnotesize (a)}
\end{minipage}\hfill
\begin{minipage}[t]{0.63\textwidth}
    \centering
    \includegraphics[width=\linewidth]{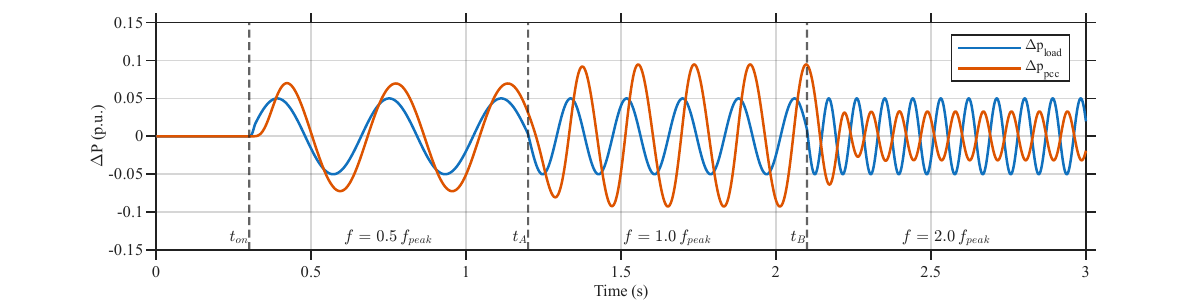}\\[-1mm]
    {\footnotesize (b)}
\end{minipage}

\caption{(a) POA factor as a function of frequency, with a peak at 5.54~\textnormal{Hz}. (b) Time-domain POA simulation under sinusoidal server-load variations.}
\label{fig:poa_combined}
\end{figure*}

\begin{figure*}[!tb]
\centering

\begin{minipage}[t]{0.33\textwidth}
\centering
\includegraphics[width=\linewidth]{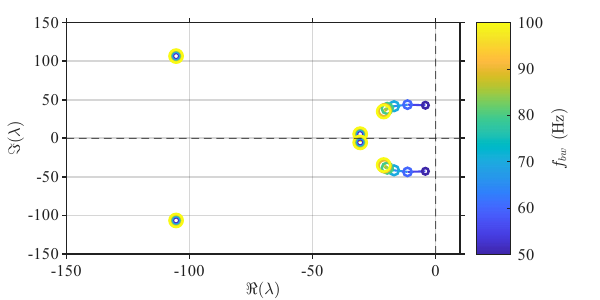}\\[-1mm]
{\footnotesize (a)}
\end{minipage}\hfill
\begin{minipage}[t]{0.33\textwidth}
\centering
\includegraphics[width=\linewidth]{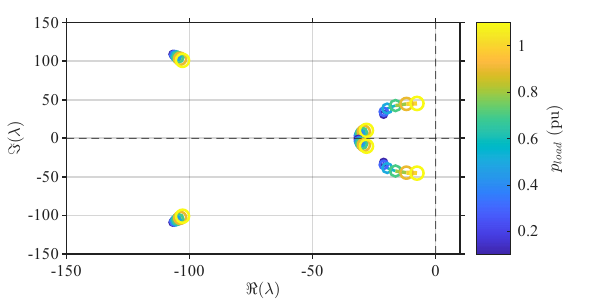}\\[-1mm]
{\footnotesize (b)}
\end{minipage}\hfill
\begin{minipage}[t]{0.33\textwidth}
\centering
\includegraphics[width=\linewidth]{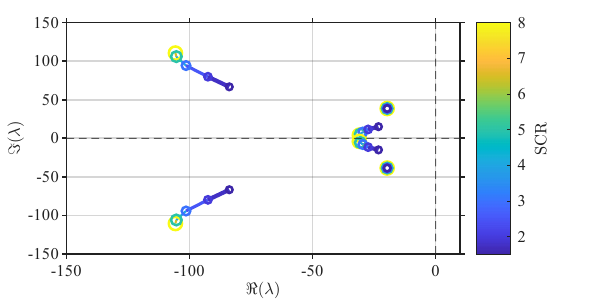}\\[-1mm]
{\footnotesize (c)}
\end{minipage}

\vspace{2mm}

\begin{minipage}[t]{0.33\textwidth}
\centering
\includegraphics[width=\linewidth]{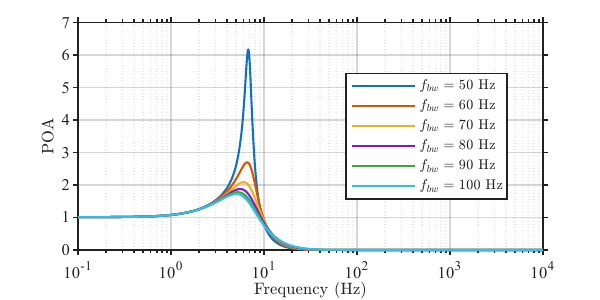}\\[-1mm]
{\footnotesize (d)}
\end{minipage}\hfill
\begin{minipage}[t]{0.33\textwidth}
\centering
\includegraphics[width=\linewidth]{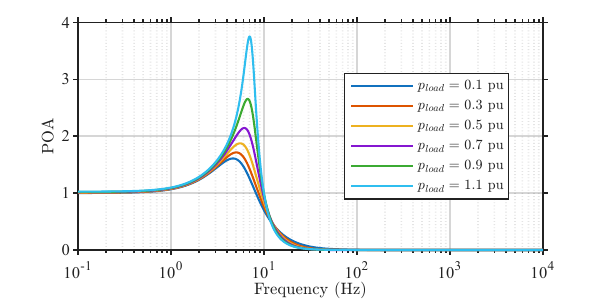}\\[-1mm]
{\footnotesize (e)}
\end{minipage}\hfill
\begin{minipage}[t]{0.33\textwidth}
\centering
\includegraphics[width=\linewidth]{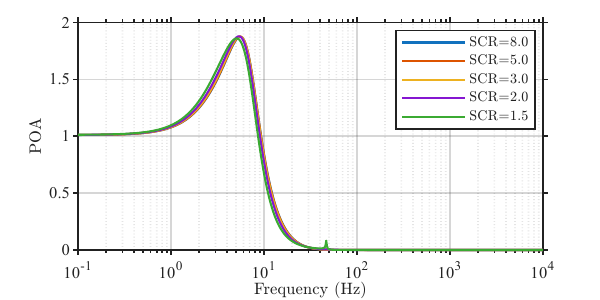}\\[-1mm]
{\footnotesize (f)}
\end{minipage}

\caption{Top-6 eigenvalue trajectories (top row) and corresponding POA curves (bottom row) under varying system conditions: (a),(d) VSI voltage-controller bandwidth; (b),(e) load level; and (c),(f) short-circuit ratio (SCR).}
\label{fig:eigen_poa_2x3}
\end{figure*}

The eigenvalue map with the corresponding (absolute) mode residues \(|\bm R_k|\) of the SDCIB case with default parameters is demonstrated in Fig. \ref{fig:eigen}. Applying participation-factor analysis, we identify the most influential state variables and their associated control loops contributing to each mode. Table~\ref{tab:pf_all_modes_top3_compact_1col} reveals a clear attribution of modal dynamics across the data-center power-delivery chain, especially:
\begin{itemize}
    \item Modes~1--2 are governed by the PSU  voltage-loop states $(v^{\textnormal{psu}},\,\xi^{\textnormal{psu}})$ together with the VSI voltage-loop integrator $\xi_{u}^{\textnormal{vsi}}$, showing a coupled PSU--VSI voltage-loop resonance
     \item Modes~3--4 are  governed by the UPS  DC-link voltage $v_{\textnormal{dc}}^{\textnormal{ups}}$ and the AFE DC-voltage-loop integrator $\xi_{\textnormal{dc}}^{\textnormal{afe}}$, reflecting an AFE--DC-link interaction.
    \item Modes~5--6 are governed by the PLL states $(\theta_{\textnormal{pll}},\,\epsilon_{\textnormal{pll}},\,v_{q}^{\textnormal{pll}})$, suggesting possible resonance caused by PLL.
\end{itemize}
The remaining modes are mainly shaped by faster inner-loop control and filter dynamics (e.g., AFE and VSI current-loop integrators and filter states), and therefore are less likely to drive the low-frequency resonance. The POA factor is shown in Fig.~\ref{fig:poa_combined}.(a). It is observed that POA curve reaches maximum at $f_\textnormal{peak}=5.54~\textnormal{Hz}$. This peak is consistent with the mode residues shown in  Fig.~\ref{fig:eigen} as well as the modal and participation results in Table~\ref{tab:pf_all_modes_top3_compact_1col}.
Specifically, the POA peak frequency lies in the vicinity of the lightly damped conjugate pair Modes~1--2 (with $f= 6.15~\textnormal{Hz}$) with large mode residue. 
This indicates that the POA peak is primarily driven by this PSU--VSI coupled voltage-regulation mode. Note that the POA peak frequency does not need to coincide exactly with the modal frequency because it can be shifted by the superposition of nearby modes. We also observe that the POA factor exhibits a low-pass characteristic. This is because the high-frequency oscillatory modes are either weakly excited by load variations or barely observable at the grid, which is consistent with the small modal residues of high-frequency modes shown in Fig.~\ref{fig:eigen}. A visualization of the POA phenomenon is given in Fig. \ref{fig:poa_combined}.(b) using time-domain simulation. The figure shows the power deviations from the equilibrium point
under sinusoidal load perturbations with amplitude $0.05$~p.u.
The dashed lines mark the excitation turn-on time $t_{\mathrm{on}}$
and the frequency switching instants $t_A$ and $t_B$.
The excitation frequencies are chosen based on the POA result: the injected sinusoid
uses piecewise-constant frequencies $0.5f_{\mathrm{peak}}$ for $t\in[t_{\mathrm{on}},t_A)$,
$1.0f_{\mathrm{peak}}$ for $t\in[t_A,t_B)$, and $2.0f_{\mathrm{peak}}$ for $t\ge t_B$. The results are consistent with Fig.~\ref{fig:poa_combined}.(a) as the PCC power fluctuation is noticeably amplified when the excitation frequency is $f_{\mathrm{peak}}$, while the response is less pronounced away from the peak frequency.
\subsubsection{Sensitivity Analysis}
Since real-world data centers may operate under varying load levels and control parameters, it is important to understand how oscillatory dynamics change with respect to setpoint and parameter shifts. In this work, we investigate the following root causes of oscillations by tracking the trajectories of the dominant oscillatory modes and the corresponding POA curves:
\begin{itemize}
    \item \emph{VSI voltage-loop bandwidth:} According to Table \ref{tab:pf_all_modes_top3_compact_1col}, the coupling of VSI-PSU voltage-loop controllers is the main cause of dominant oscillatory modes 1--2. By successively varying the bandwidth  of the VSI voltage-loop controller, the trajectories of the top-6 oscillatory modes with highest mode residue are shown in Fig.~\ref{fig:eigen_poa_2x3}.a. Accordingly, the POA curves under changing $f_\textnormal{bw}$ are shown in Fig.~\ref{fig:eigen_poa_2x3}.d. Once the voltage-loop bandwidth separation of VSI and PSU is lost (i.e. the bandwidth $f_\textnormal{bw}$ of  VSI voltage-loop decreases), both loops act on the same dynamics at the same time. This creates a delayed “over-correction” cycle, which reduces damping, amplifies resonance and can even lead to unstable modes. 
    \item \emph{Load level:} As the load level $p_\textnormal{load}$ is gradually increased from $0.2$ to $1.0$~p.u., modes~1--2 move quickly toward the imaginary axis as shown in the Fig. \ref{fig:eigen_poa_2x3}.(b), which leads to reduced damping and thereby results in a noticeably larger peak in the POA curves as shown in the  Fig. \ref{fig:eigen_poa_2x3}.(e). Given that the PSU and VSI  voltage loops are coupled through the  AC-bus current, increasing the load raises the steady-state current, thereby strengthening the coupling effects.
    \item \emph{Grid strength:} Under the per-unit system, the short-circuit ratio (SCR) is defined as $\mathrm{SCR}=1/{|Z_{\infty}|=\sqrt{R_{\infty}^2+X_{\infty}^2}}$. The SCR is a commonly-used indicator of grid strength.  When reducing the SCR (i.e., reducing grid strength) with a fixed $R_{\infty}$--$X_{\infty}$ ratio in the Table \ref{tab:sdcib_params}, the eigenvalues of mode 3-6 move towards the imaginary axis as shown in the Fig. \ref{fig:eigen_poa_2x3}.(c). However, as shown in Fig.~\ref{fig:eigen_poa_2x3}.(f), the POA curves remain almost unchanged, primarily because high-residue modes 1--2 are weakly sensitive to the SCR, whereas hightly-sensitive modes 5--6 have relatively low mode residues.
\end{itemize}

\begin{figure*}[!tb]
\centering
\begin{minipage}[t]{0.5\textwidth}
    \centering
    \includegraphics[width=\linewidth]{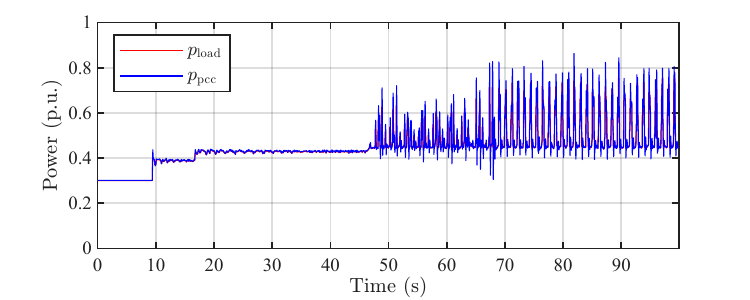}\\[-1mm]
    {\footnotesize (a)}
\end{minipage}\hfill
\begin{minipage}[t]{0.5\textwidth}
    \centering
    \includegraphics[width=\linewidth]{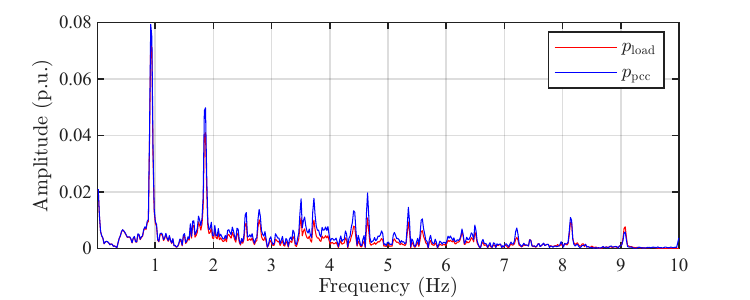}\\[-1mm]
    {\footnotesize (b)}
\end{minipage}
\caption{Realistic GPU-induced PCC power profile: (a) time-domain simulation. (b) single-sided frequency spectrum from FFT for $t\in[50,100]$s.}
\label{fig:ai_load_fft}
\end{figure*}

\subsubsection{Time-domain Simulation under Realistic GPU load} To study how AI training load profiles propagate through the power stages of a data center, a realistic GPU-cluster workload trace with 100-ms resolution from the MIT Supercloud dataset \cite{mitsupercloud_dataset}, which is further processed in \cite{aghadinuno2026ai}, is used as the scaled server-load profile $p_\textnormal{load}(t)$. The time-domain results in Fig. \ref{fig:ai_load_fft}.(a) show that the server-side AI-training load fluctuation is transmitted through the data-center power-delivery stages to the PCC. Although the PCC power generally follows the load-power variation, more pronounced oscillatory peaks are observed at the PCC in the later stage ($t\in[50,100]$s), indicating that the conversion chain can amplify certain dynamic components rather than merely smoothing them. This is further confirmed by the FFT results in Fig.~\ref{fig:ai_load_fft}(b), where the dominant spectral components (in the range of $1$--$2~\textnormal{Hz}$) of $p_{\mathrm{pcc}}$ closely match those of $p_{\mathrm{load}}$. However, larger amplitudes are observed at several frequencies, with the POA factor in Fig.~\ref{fig:poa_combined}(a) exceeding unity, (particularly in the vicinity of $f_\textnormal{peak}=5.54  \textnormal{ Hz}$). This indicates the frequency-selective propagation and amplification of workload-induced power oscillations.

\subsection{Modified 3-Machine 9-Bus System} 
\begin{figure}[!tb]
    \centering
\includegraphics[width=0.9\linewidth]{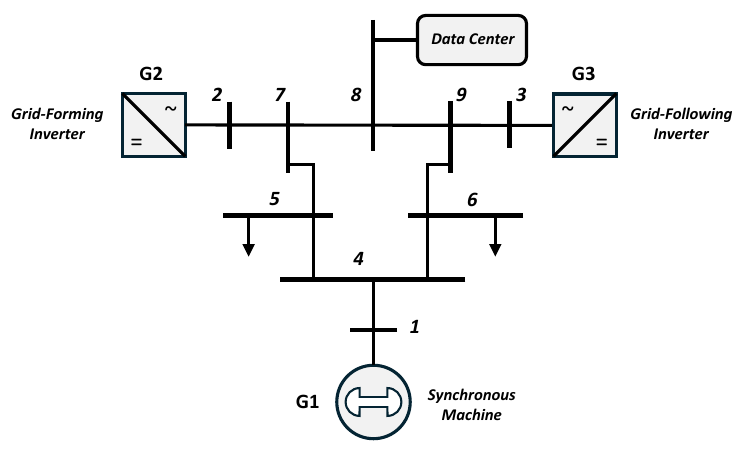}
    \caption{Modified 3-machine 9-bus system.}
    \label{fig:wscc9}
\end{figure}

To validate the grid-integration capability of the proposed model and to investigate the dynamic coupling between the data center and external grids, especially inverter-based resources, a modified 3-machine 9-bus system is established, as shown in Fig.~\ref{fig:wscc9}. In this system, $G1$ is represented by a synchronous machine (SM) using a fourth-order generator model (i.e., a two-axis transient model) equipped with an IEEE DC1A exciter and a TGOV1 turbine governor (see the detailed 9th-order formulation in \eqref{app:eqsm} of the Supplementary Material). $G2$ is represented by a grid-forming inverter (GFM) model (see the detailed 13th-order formulation in \eqref{app:eqgfm} of the Supplementary Material), and $G3$ is represented by a grid-following inverter (GFL) model (see the detailed 15th-order formulation in \eqref{app:eqgfl} of the Supplementary Material). Their parameters are summarized in Table~\ref{tab:all_params} of the Supplementary Material.  A data center represented by \eqref{eq:afe}--\eqref{eq:dcdc_qss_xi_state} using parameter setting in Table.\ref{tab:sdcib_params} is linked to Bus 8 with unity power factor.

\begin{figure*}[!tb]
     \centering
    \includegraphics[width=\linewidth]{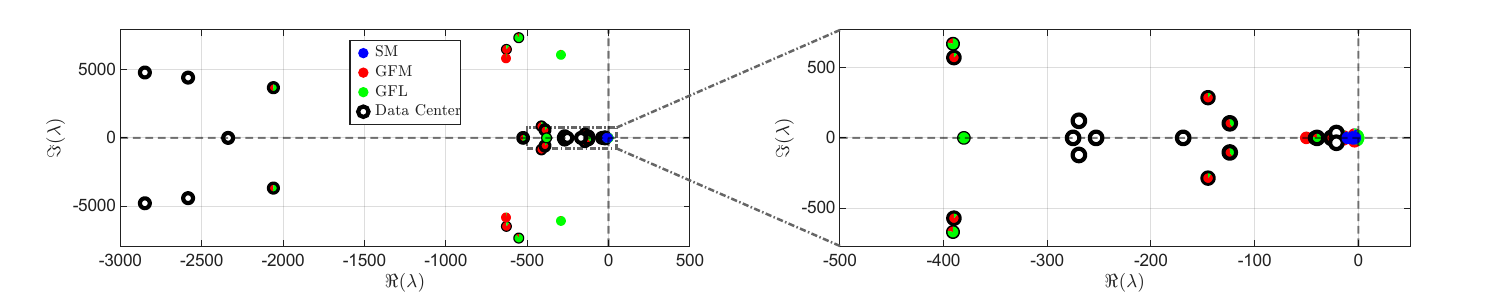}
     \caption{Eigenvalue map with coupling visualization of the  modified 3-machine 9-bus system}
     \label{fig:eigen9bus}
 \end{figure*} 

\begin{figure}[!tb]
     \centering
    \includegraphics[width=0.8\linewidth]{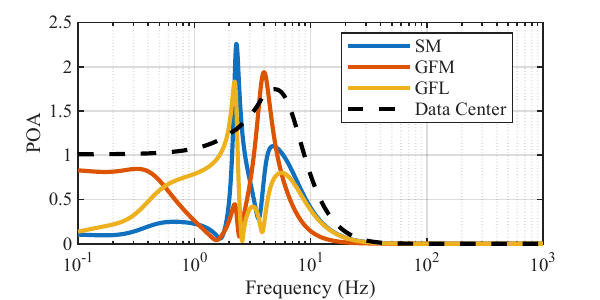}
     \caption{POA from  server load to SM, GFM, GFL, and data center port}
     \label{fig:poa9bus}
 \end{figure} 

\begin{figure}[!tb]
     \centering
    \includegraphics[width=0.9\linewidth]{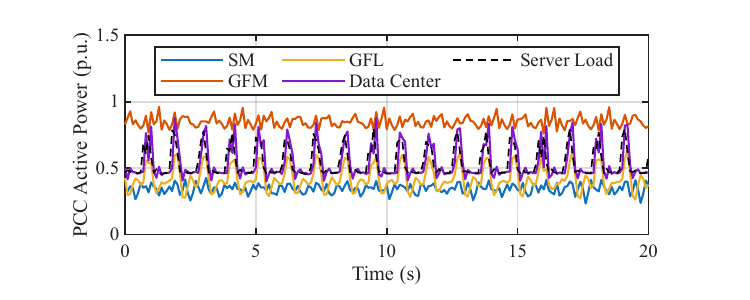}
     \caption{Realistic GPU-induced PCC active power of the multi-machine case}
     \label{fig:pport}
 \end{figure} 

\subsubsection{Load-Driven Oscillation Analysis}

The eigenvalue map of the modified 3-machine 9-bus system is shown in Fig.~\ref{fig:eigen9bus}. Each marker indicates modal composition based on the participation of the SM, GFM, and GFL states, represented by blue, red, and green sectors, respectively. The influence of the data-center dynamics is highlighted by the black outer ring, whose thickness reflects the participation of the data-center states. From the zoomed-in view, three frequency ranges can be observed. For the dominant slow modes closest to the origin, the markers cluster near the real axis and show mixed participation from the SM, GFM, and GFL states, whereas the black outer rings are generally weak or absent, indicating that the data-center states have only limited participation in these modes. Therefore, the modes closest to the stability boundary are governed primarily by the grid-side electromechanical and inverter-control dynamics. As the oscillation frequency increases within the zoomed-in range, the markers exhibit more visible black outer rings, indicating stronger coupling with the data-center power-delivery dynamics. In this range, the modal composition also shifts toward stronger GFM/GFL participation and comparatively weaker SM contribution. Overall, the dominant modal structure is mainly determined by the SM, GFM, and GFL dynamics, while the data-center subsystem primarily modifies existing modes rather than introducing new poorly damped oscillatory modes. The POA factor analysis is shown in Fig.~\ref{fig:poa9bus}. Based on the state-space model of the interconnected system, the POA framework previously developed for the single-machine case is extended to a multi-machine, multi-port setting to quantify oscillation transfer from the server load to the SM, GFM, GFL, and data-center ports. The results reveal a pronounced amplification band around 2--6 Hz, where load-induced oscillations are significantly amplified by the coupled grid--converter dynamics. The SM port shows the strongest response in the low-frequency range, indicating excitation of electromechanical modes, while the GFM and GFL ports become more responsive at intermediate frequencies due to inverter-control dynamics. In contrast, the data-center port exhibits a broader amplification profile, indicating that the data-center power-delivery chain acts as the primary disturbance source. For frequencies above approximately 20 Hz, the amplification rapidly decreases, showing that high-frequency workload fluctuations are effectively filtered by the converter stages.

\subsubsection{Time-domain Simulation under Realistic GPU load} 
The same GPU-cluster workload trace shown in Fig.~\ref{fig:ai_load_fft} is adopted in this case to investigate how the grid-side generation units, i.e., the SM, GFM, and GFL, respond to server-load oscillations. The time-domain simulation results corresponding to the last 10~s of the workload trace in Fig.~\ref{fig:ai_load_fft}. (a) are shown in Fig.~\ref{fig:pport}. From Fig.~\ref{fig:pport}, all grid-side generation units exhibit oscillatory active-power responses to the server-load variations, indicating that the workload-induced power fluctuations are transmitted through the data-center power-delivery stages into the external grid. Among the three generation units, the GFL exhibits the largest oscillation amplitude, while the SM  and GFM responses are comparatively smaller. This suggests that the GFL is more sensitive to the workload-driven power oscillations in this case.

\section{Conclusions}
\label{con}
This paper derived a component-informed dynamic model of the data-center power-delivery chain for power-system oscillation studies. The proposed model preserves the key converter dynamics and control interactions while providing a time-invariant positive-sequence representation suitable for phasor-domain simulation and small-signal analysis. The results on the SMIB case showed that intrinsic oscillatory modes exist within the data-center power path, among which the coupled VSI--PSU voltage-control mode is the main source of load-driven oscillation amplification. The proposed POA framework further revealed how server-load fluctuations are selectively amplified and propagated to the PCC. Case studies on the modified 3-machine 9-bus system showed that workload-induced power fluctuations can also propagate into external grids and interact with SM, GFM and GFL. Overall, the proposed framework provides a transparent tool for understanding and mitigating oscillation risks of grid-connected data centers. The results also provide practical insights for grid integration of large data centers. In particular, the internal power-delivery chain should be explicitly considered in interconnection studies rather than being represented as a static load. Coordinated controller bandwidth design across the data-center converters is also important to avoid poorly damped resonance modes. Moreover, AI workload fluctuations may act as sustained disturbances capable of exciting grid oscillations, especially in converter-dominated systems.

Future research will scale the proposed framework to multi-data-center systems and integrate high-fidelity workload models. Key areas of exploration include coordinated oscillation-mitigation strategies and the interaction between AI scheduling and grid stability. Furthermore, we will account for transmission line dynamics to evaluate their impact on the  eigenvalue characteristics.


%





\ifCLASSOPTIONcaptionsoff
  \newpage
\fi



%


\clearpage
\onecolumn %

\setcounter{section}{0}
\setcounter{equation}{0}
\setcounter{figure}{0}
\setcounter{table}{0}
\setcounter{page}{1} %

\renewcommand{\thesection}{S\arabic{section}}
\renewcommand{\theequation}{S\arabic{equation}}
\renewcommand{\thefigure}{S\arabic{figure}}
\renewcommand{\thetable}{S\arabic{table}}

\begin{center}
    \Large \textbf{Supplementary Material for \\``Dynamic Modeling of Data-Center Power Delivery for Power System Resonance Analysis"}
\end{center}

\providecommand{\d}{\mathrm{d}}
\section{Derivation of Positive-Sequence Dynamic Equivalent}
\subsection{PSU Array}
\label{app:psu}
The single-phase PSU control block in Fig.~\ref{fig:psu} is first written in DAE form and then stacked across three phases in the $abc$ frame, yielding the PSU-array model in \eqref{eq:pfc_abc_dcside}.
\begin{subequations}
\label{eq:pfc_abc_dcside}
\allowdisplaybreaks
\setlength{\jot}{2pt}
\thinmuskip=2mu
\medmuskip=2mu
\thickmuskip=2mu
\begin{align}
\shortintertext{\textit{Differential Equations}}
\frac{\ell_\textnormal{psu}}{\omega_b}\,
\frac{\d \bm i_{abc}^{\textnormal{rec}}}{\d t}
&=
\bm v_{abc}^{\textnormal{rec}}
-
\bigl(\bm 1-\bm d_{abc}\bigr)\odot \bm v_{abc}^{\textnormal{psu}}
-
r_\textnormal{psu}\,\bm i_{abc}^{\textnormal{rec}},
\\
\frac{c_\textnormal{psu}}{\omega_b}\,
\frac{\d \bm v_{abc}^{\textnormal{psu}}}{\d t}
&=
\bigl(\bm 1-\bm d_{abc}\bigr)\odot \bm i_{abc}^{\textnormal{rec}}
-
\bm i_{abc}^{\textnormal{psu}},
\\
\frac{\d \bm \xi_{abc}^{\textnormal{psu}}}{\d t}
&=
v_{\textnormal{psu}}^{\textnormal{ref}}\,\bm 1
-
\bm v_{abc}^{\textnormal{psu}},
\\
\frac{\d \bm \gamma_{abc}^{\textnormal{psu}}}{\d t}
&=
\bm i_{abc}^{\textnormal{rec,ref}}
-
\bm i_{abc}^{\textnormal{rec}}.
\\[2pt]
\shortintertext{\textit{Algebraic Equations}}
\bm g_{\textnormal{eq},abc}
&=
k_p^{\textnormal{v,psu}}
\bigl(
v_{\textnormal{psu}}^{\textnormal{ref}}\,\bm 1
-
\bm v_{abc}^{\textnormal{psu}}
\bigr)
+
k_i^{\textnormal{v,psu}}\,
\bm \xi_{abc}^{\textnormal{psu}},
\\
\bm i_{abc}^{\textnormal{rec,ref}}
&=
\bm g_{\textnormal{eq},abc}\odot \bm v_{abc}^{\textnormal{rec}},
\\
\bm d_{abc}
&=
k_p^{\textnormal{c,psu}}
\bigl(
\bm i_{abc}^{\textnormal{rec,ref}}
-
\bm i_{abc}^{\textnormal{rec}}
\bigr)
+
k_i^{\textnormal{c,psu}}\,
\bm \gamma_{abc}^{\textnormal{psu}},
\\
0 &\le d_{\phi} \le 1,\; \phi\in\{a,b,c\}.
\end{align}
\end{subequations}
\textit{Additional Symbols:}
For arbitrary three-phase quantity, $\bm x_{abc}\triangleq [x_a \,\, x_b \,\, x_c]^\top$; $\odot$ and $\oslash$ denote element-wise multiplication and division, respectively. In \eqref{eq:pfc_abc_dcside}, $\bm\gamma_{abc}^{\textnormal{psu}}$ is the inner-loop integrator state, $\bm i_{abc}^{\textnormal{rec,ref}}$ is the rectifier-current reference, and $\bm d_{abc}$ is the duty-ratio command. $\ell_\textnormal{psu}$ is the per-phase PFC inductance, and $(k_p^{\textnormal{c,psu}},k_i^{\textnormal{c,psu}})$ are the inner current-loop PI gains. 
Assuming a clear time-scale separation between the outer voltage and inner current loop, we apply a QSS approximation to the inner loop by enforcing $\bm i_{abc}^{\textnormal{rec,ref}}=\bm i_{abc}^{\textnormal{rec}}$ and $\d \bm i_{abc}^{\textnormal{rec}}/\d t\approx \bm 0$. Eliminating $\bm i_{abc}^{\textnormal{rec,ref}}$, $\bm\gamma_{abc}^{\textnormal{psu}}$, and $\bm d_{abc}$ gives:
\begin{subequations}
\label{eq:psu_abc_recpsu_qss}
\allowdisplaybreaks
\setlength{\jot}{2pt}
\thinmuskip=2mu
\medmuskip=2mu
\thickmuskip=2mu
\begin{align}
\shortintertext{\textit{Differential Equations}}
\frac{c_\textnormal{psu}}{\omega_b}\,
\frac{\d \bm v_{abc}^{\textnormal{psu}}}{\d t}
&=
\Big(
\bm v_{abc}^{\textnormal{rec}}\odot \bm i_{abc}^{\textnormal{rec}}
-
r_\textnormal{psu}\,\bm i_{abc}^{\textnormal{rec}}\odot \bm i_{abc}^{\textnormal{rec}}
\Big)\oslash \bm v_{abc}^{\textnormal{psu}}
-
\bm i_{abc}^{\textnormal{psu}},
\\
\frac{\d \bm \xi_{abc}^{\textnormal{psu}}}{\d t}
&=
v_{\textnormal{psu}}^{\textnormal{ref}}\bm 1
-
\bm v_{abc}^{\textnormal{psu}}.
\\[2pt]
\shortintertext{\textit{Algebraic Equations}}
\bm g_{\textnormal{eq},abc}
&=
k_p^{\textnormal{v,psu}}
\bigl(
v_{\textnormal{psu}}^{\textnormal{ref}}\,\bm 1
-
\bm v_{abc}^{\textnormal{psu}}
\bigr)
+
k_i^{\textnormal{v,psu}}\,
\bm \xi_{abc}^{\textnormal{psu}},
\\
\bm i_{abc}^{\textnormal{rec}}
&=
\bm g_{\textnormal{eq},abc}\odot \bm v_{abc}^{\textnormal{rec}}.
\end{align}
\end{subequations}
To rewrite the reduced model on the AC side (VSI side), introduce the phase-wise sign vector
\[
\bm s_{abc}\triangleq
\begin{bmatrix}
\mathrm{sgn}\!\bigl(v_a^{\textnormal{vsi}}\bigr) &
\mathrm{sgn}\!\bigl(v_b^{\textnormal{vsi}}\bigr) &
\mathrm{sgn}\!\bigl(v_c^{\textnormal{vsi}}\bigr)
\end{bmatrix}^{\top},
\]
where $\mathrm{sgn}(\cdot)\in\{-1,0,1\}$ is the scalar sign function. This sign mapping captures the diode-bridge conduction polarity, keeping rectified-side variables positive. Using $\bm v_{abc}^{\textnormal{vsi}}=\bm s_{abc}\odot \bm v_{abc}^{\textnormal{rec}}$ and $\bm i_{abc}^{\textnormal{vsi}}=\bm s_{abc}\odot \bm i_{abc}^{\textnormal{rec}}$, together with $\bm s_{abc}\odot \bm s_{abc}=\bm 1$ (away from zero crossings), the capacitor dynamics become
\[
\frac{c_\textnormal{psu}}{\omega_b}\,
\frac{\d \bm v_{abc}^{\textnormal{psu}}}{\d t}
=
\Big(
\bm v_{abc}^{\textnormal{vsi}}\odot \bm i_{abc}^{\textnormal{vsi}}
-
r_\textnormal{psu}\,\bm i_{abc}^{\textnormal{vsi}}\odot \bm i_{abc}^{\textnormal{vsi}}
\Big)\oslash \bm v_{abc}^{\textnormal{psu}}
-\bm i_{abc}^{\textnormal{psu}},
\]
and the conductance-form current relation is preserved:
\[
\bm i_{abc}^{\textnormal{vsi}}
=
\bm g_{\textnormal{eq},abc}\odot \bm v_{abc}^{\textnormal{vsi}}.
\]
Next, under the common-DC approximation (neglecting the DC-side $2\omega$ ripple and phase variations), we have $\bm v_{abc}^{\textnormal{psu}}\approx v^{\textnormal{psu}}\bm 1$,
$\bm \xi_{abc}^{\textnormal{psu}}\approx \xi^{\textnormal{psu}}\bm 1$, and $\bm g_{\textnormal{eq},abc}\approx g_\textnormal{eq}\bm 1$,
yielding
\[
g_\textnormal{eq}
=
k_p^\textnormal{v,psu}\bigl(v_{\textnormal{psu}}^{\textnormal{ref}}-v^{\textnormal{psu}}\bigr)
+
k_i^\textnormal{v,psu}\,\xi^{\textnormal{psu}},
\qquad
\bm i_{abc}^{\textnormal{vsi}}
=
g_\textnormal{eq}\bm v_{abc}^{\textnormal{vsi}}.
\]
Summing the three phase-capacitor equations gives the aggregated DC-port dynamics:
\[
\begin{aligned}
\frac{c_\textnormal{psu}}{\omega_b}\,
\frac{\d v^{\textnormal{psu}}}{\d t}
&=
\frac{1}{3v^{\textnormal{psu}}}\,
\bm 1^\top\!\left(
\bm v_{abc}^{\textnormal{vsi}}\odot \bm i_{abc}^{\textnormal{vsi}}
-
r_\textnormal{psu}\,\bm i_{abc}^{\textnormal{vsi}}\odot \bm i_{abc}^{\textnormal{vsi}}
\right)
-
i^{\textnormal{psu}}
\\
&=
\frac{1}{3v^{\textnormal{psu}}}\left(
\left(\bm v_{abc}^{\textnormal{vsi}}\right)^\top \bm i_{abc}^{\textnormal{vsi}}
-
r_\textnormal{psu}\,\|\bm i_{abc}^{\textnormal{vsi}}\|_2^2
\right)
-
i^{\textnormal{psu}}.
\end{aligned}
\]
Finally, we express the AC-side variables in the $uv$ frame via the \emph{power-invariant} Clarke/Park transformation with angle $\theta_\textnormal{vsi}$. Under the balanced three-phase assumption, we obtained $\bm i_{uv}^{\textnormal{vsi}}
=
g_\textnormal{eq}\bm v_{uv}^{\textnormal{vsi}}$, such that
\[
\left(\bm v_{abc}^{\textnormal{vsi}}\right)^\top \bm i_{abc}^{\textnormal{vsi}}
=
g_\textnormal{eq}\,\|\bm v_{uv}^{\textnormal{vsi}}\|_2^2,\qquad
\|\bm i_{abc}^{\textnormal{vsi}}\|_2^2
=
g_\textnormal{eq}^2\,\|\bm v_{uv}^{\textnormal{vsi}}\|_2^2.
\]
Hence,
\[
\left(
\left(\bm v_{abc}^{\textnormal{vsi}}\right)^\top \bm i_{abc}^{\textnormal{vsi}}
-
r_\textnormal{psu}\,\|\bm i_{abc}^{\textnormal{vsi}}\|_2^2
\right)
=
\Bigl(g_\textnormal{eq}-r_\textnormal{psu}g_\textnormal{eq}^2\Bigr)\,
\|\bm v_{uv}^{\textnormal{vsi}}\|_2^2.
\]
Substituting the above relations into the aggregated DC-port dynamics yields the compact common-DC PSU-array model in \eqref{eq:psu_uv_qss_xi_state}. 

\subsection{Downstream DC--DC Converter and Load}
\label{app:eq}
The downstream DC--DC/load equivalent in Fig.~\ref{fig:dcdc_eq} is first written in the DAE form \eqref{eq:dcdc_buck_equiv_scalar_currloop}.
\begin{subequations}
\label{eq:dcdc_buck_equiv_scalar_currloop}
\allowdisplaybreaks
\setlength{\jot}{2pt}
\thinmuskip=2mu
\medmuskip=2mu
\thickmuskip=2mu
\begin{align}
\shortintertext{\textit{Differential Equations}}
\frac{\ell_\textnormal{eq}}{\omega_b}\,
\frac{\d i^\textnormal{eq}}{\d t}
&=
d_\textnormal{eq}\, v^\textnormal{psu}
-
v_o^\textnormal{eq},
\\[2pt]
\frac{c_\textnormal{eq}}{\omega_b}\,
\frac{\d v_o^\textnormal{eq}}{\d t}
&=
i^\textnormal{eq}
-
g_\textnormal{load}\, v_o^\textnormal{eq},
\\[2pt]
\frac{\d \xi^\textnormal{eq}}{\d t}
&=
v_o^{\textnormal{eq,ref}}
-
v_o^\textnormal{eq},
\\[2pt]
\frac{\d \gamma^\textnormal{eq}}{\d t}
&=
i^{\textnormal{eq,ref}}
-
i^\textnormal{eq}.
\\[4pt]
\shortintertext{\textit{Algebraic Equations}}
g_\textnormal{load}
&=
\frac{p_\textnormal{load}}{3\left(v_o^{\textnormal{eq,ref}}\right)^2},
\\[2pt]
i^{\textnormal{eq,ref}}
&=
k_p^\textnormal{v,eq}
\bigl(
v_o^{\textnormal{eq,ref}}
-
v_o^\textnormal{eq}
\bigr)
+
k_i^\textnormal{v,eq}\,
\xi^\textnormal{eq},
\\[2pt]
d_\textnormal{eq}
&=
k_p^\textnormal{c,eq}
\bigl(
i^{\textnormal{eq,ref}}
-
i^\textnormal{eq}
\bigr)
+
k_i^\textnormal{c,eq}\,
\gamma^\textnormal{eq},
\\[2pt]
i^\textnormal{psu}
&=
d_\textnormal{eq}\, i^\textnormal{eq},\\
0 &\le d_{\textnormal{eq}} \le 1.
\end{align}
\end{subequations}
\textit{Additional Symbols:}
$\ell_\textnormal{eq}$ are the equivalent inductance of the downstream DC--DC stage; $(k_p^\textnormal{c,eq},k_i^\textnormal{c,eq})$ are the inner current-loop PI gains; $i^{\textnormal{eq,ref}}$ is the current reference generated by the outer voltage loop; $d_\textnormal{eq}$ is the duty-ratio command; $\gamma^\textnormal{eq}$ is the inner-loop integrator state; $i^\textnormal{psu}$ is the input current drawn from the upstream PSU DC port; and $v^\textnormal{psu}$ is the PSU DC-port voltage. Under the common-DC approximation (i.e., neglecting the $2\omega$ ripple on the PSU DC port), $i^\textnormal{psu}\approx i_\phi^\textnormal{psu}$ and $v^\textnormal{psu}\approx v_\phi^\textnormal{psu}$.
Assuming a clear time-scale separation between the outer voltage loop and the inner current loop, a QSS approximation is applied to the inner current loop by enforcing $i^{\textnormal{eq}}=i^{\textnormal{eq,ref}}$ and $\d i^{\textnormal{eq}}/\d t \approx 0$. This eliminates the inner-loop variables $(i^{\textnormal{eq,ref}},\gamma^\textnormal{eq},d_\textnormal{eq})$ and yields the reduced DAE model in \eqref{eq:dcdc_qss_xi_state}.

\subsection{Extra Parameter Setup of The Full-Order Model}

\begin{table}[!hb]
\centering
\caption{Extra Default Parameter Setup for the Full-Order Model}
\label{tab:extra_params}
\renewcommand{\arraystretch}{1}
\setlength{\tabcolsep}{2pt}
\begin{tabular}{llp{6cm}}
\toprule
\textbf{Symbol} & \textbf{Value} & \textbf{Description} \\
\midrule
$\ell_\textnormal{psu}$ & $0.05~\textnormal{p.u.}$ & PSU AC-side inductance   \\
$k_p^\textnormal{c,psu}$, $k_i^\textnormal{c,psu}$ & $ 1.6617$, $5236$ & PSU current-PI ($f_{\mathrm{bw}}=1000~\textnormal{Hz},\,\zeta=1.0$) \\

$\ell_\textnormal{eq}$ & $0.05~\textnormal{p.u.}$ & DC-DC load inductance   \\

$k_p^\textnormal{c,eq}$, $k_i^\textnormal{c,eq}$ & $ 1.6617$, $5236$ & DC-DC load current-PI ($f_{\mathrm{bw}}=1000~\textnormal{Hz},\,\zeta=1.0$) \\

\bottomrule
\end{tabular}
\end{table}

\section{Bandwidth-Based PI Controller Tuning Rule}
\label{app:tuning}
The PI controller gains in Table \ref{tab:sdcib_params} are obtained from a bandwidth-based tuning rule with target bandwidth $f_{\mathrm{bw}}$ and damping ratio $\zeta$, by matching each local PI-controlled loop to the standard second-order form
$ s^2 + 2\zeta\omega_n s + \omega_n^2 $, where $\omega_n = 2\pi f_{\mathrm{bw}}$. The tuning rules are: 
\begin{itemize}
    \item \textit{Voltage loops:} $k_p = 2\zeta\omega_n C/\omega_b$ and $k_i = \omega_n^2 C/\omega_b$, where $C$ is the effective capacitance.

    \item \textit{Current loops:} $k_p = 2\zeta\omega_n L/\omega_b-R$ and $k_i = \omega_n^2 L/\omega_b$, where $L$ and $R$ are the loop inductance and resistance.

    \item \textit{PLL:} $k_p=2\zeta\omega_n/\omega_b$ and $k_i=\omega_n^2/\omega_b$.
\end{itemize}

This tuning strategy enforces the intended time-scale separation among cascaded loops (i.e., inner current loops are tuned faster than outer voltage loops) and provides a consistent and reproducible way to set all PI gains.

\section{The Modified WSCC 3-Machine 9-Bus System}
\subsection{Mathematical Formulation}

\subsubsection{Synchronous Machine with Exciter and Governor}

\begin{subequations}
\label{app:eqsm}
{\allowdisplaybreaks
\begin{align}
\textit{Differential Eqs.} \notag \\[0.3em]
\frac{1}{\omega_b}\frac{\mathrm{d}\delta}{\mathrm{d}t}
&=
\omega-\omega_s
\label{eq:sm_delta}
\\
2H\frac{\mathrm{d}\omega}{\mathrm{d}t}
&=
\tau_m-\tau_e-D(\omega-\omega_s)
\label{eq:sm_omega}
\\
T_{d0}^{\prime}\frac{\mathrm{d}e_q^{\prime}}{\mathrm{d}t}
&=
-e_q^{\prime}
-(x_d-x_d^{\prime})i_d
+e_{fd}
\label{eq:sm_eqp}
\\
T_{q0}^{\prime}\frac{\mathrm{d}e_d^{\prime}}{\mathrm{d}t}
&=
-e_d^{\prime}
+(x_q-x_q^{\prime})i_q
\label{eq:sm_edp}
\\
T_e\frac{\mathrm{d}e_{fd}}{\mathrm{d}t}
&=
-\bigl(k_e+s_e(e_{fd})\bigr)e_{fd}
+v_r
\label{eq:sm_efd}
\\
T_f\frac{\mathrm{d}v_f}{\mathrm{d}t}
&=
-v_f
+\frac{k_f}{T_e}v_r
-\frac{k_f}{T_e}\bigl(k_e+s_e(e_{fd})\bigr)e_{fd}
\label{eq:sm_vf}
\\
T_a\frac{\mathrm{d}v_r}{\mathrm{d}t}
&=
-v_r
+k_a(v_{\mathrm{ref}}-v_f-v_t)
\label{eq:sm_vr}
\\
T_{sv}\frac{\mathrm{d}p_{sv}}{\mathrm{d}t}
&=
-p_{sv}
+p_{\mathrm{ref}}
-\frac{1}{r}(\omega-\omega_s)
\label{eq:sm_psv}
\\
T_{ch}\frac{\mathrm{d}\tau_m}{\mathrm{d}t}
&=
-\tau_m+p_{sv}
\label{eq:sm_taum}
\\[0.3em]
\textit{Algebraic Eqs.} \notag \\[0.3em]
s_e(e_{fd})
&=
a_e \exp(b_e e_{fd})
\label{eq:sm_se}
\\
\mathbf{v}_{dq}^{\mathrm{gen}}
&=
\mathbf{T}(\delta)\,
\mathbf{v}_{ri}^{\mathrm{gen}}
\label{eq:sm_vdq}
\\
i_d
&=
\frac{e_q^{\prime}-v_q}{x_d^{\prime}}
\label{eq:sm_id}
\\
i_q
&=
\frac{v_d-e_d^{\prime}}{x_q^{\prime}}
\label{eq:sm_iq}
\\
\mathbf{i}_{ri}^{\mathrm{gen}}
&=
\mathbf{T}(\delta)^\top
\mathbf{i}_{dq}
\label{eq:sm_iri}
\\
\tau_e
&=
v_d i_d+v_q i_q
\label{eq:sm_taue}
\\
v_t
&=
\sqrt{v_d^2+v_q^2}
\label{eq:sm_vt}
\end{align}
}
\end{subequations}
\subsubsection{Grid-Forming Inverter}

\begin{subequations}
\label{app:eqgfm}
{\allowdisplaybreaks
\begin{align}
\textit{Differential Eqs.} \notag \\[0.3em]
\frac{1}{\omega_b}\frac{\mathrm{d}\theta_{\mathrm{oc}}^{\mathrm{gfm}}}{\mathrm{d}t}
&=
\omega_{\mathrm{oc}}^{\mathrm{gfm}}-\omega_s
\label{eq:gfm_theta}
\\
\frac{1}{\omega_z^{\mathrm{gfm}}}\frac{\mathrm{d}p^{\mathrm{oc,gfm}}}{\mathrm{d}t}
&=
p^{\mathrm{f,gfm}}-p^{\mathrm{oc,gfm}}
\label{eq:gfm_poc}
\\
\frac{1}{\omega_f^{\mathrm{gfm}}}\frac{\mathrm{d}q^{\mathrm{oc,gfm}}}{\mathrm{d}t}
&=
q^{\mathrm{f,gfm}}-q^{\mathrm{oc,gfm}}
\label{eq:gfm_qoc}
\\
\frac{\mathrm{d}\bm{\xi}^{\mathrm{gfm}}}{\mathrm{d}t}
&=
\bm{v}_{\mathrm{vi}}^{\mathrm{gfm}}
-
\bm{v}_{\mathrm{f}}^{\mathrm{gfm}}
\label{eq:gfm_xi}
\\
\frac{\mathrm{d}\bm{\gamma}^{\mathrm{gfm}}}{\mathrm{d}t}
&=
\bm{i}_{\mathrm{ref}}^{\mathrm{gfm}}
-
\bm{i}_{\mathrm{cv}}^{\mathrm{gfm}}
\label{eq:gfm_gamma}
\\
\frac{\ell_f^{\mathrm{gfm}}}{\omega_b}
\frac{\mathrm{d}\bm{i}_{\mathrm{cv}}^{\mathrm{gfm}}}{\mathrm{d}t}
&=
\bm{v}_{\mathrm{cv}}^{\mathrm{gfm}}
-
\bm{v}_{\mathrm{f}}^{\mathrm{gfm}}
-
r_f^{\mathrm{gfm}}\bm{i}_{\mathrm{cv}}^{\mathrm{gfm}}
-
\omega_{\mathrm{oc}}^{\mathrm{gfm}}\ell_f^{\mathrm{gfm}}
\bm{J}\bm{i}_{\mathrm{cv}}^{\mathrm{gfm}}
\label{eq:gfm_icv_dyn}
\\
\frac{c_f^{\mathrm{gfm}}}{\omega_b}
\frac{\mathrm{d}\bm{v}_{\mathrm{f}}^{\mathrm{gfm}}}{\mathrm{d}t}
&=
\bm{i}_{\mathrm{cv}}^{\mathrm{gfm}}
-
\bm{i}_{\mathrm{g}}^{\mathrm{gfm}}
-
\omega_{\mathrm{oc}}^{\mathrm{gfm}}c_f^{\mathrm{gfm}}
\bm{J}\bm{v}_{\mathrm{f}}^{\mathrm{gfm}}
\label{eq:gfm_vf_dyn}
\\
\frac{\ell_g^{\mathrm{gfm}}}{\omega_b}
\frac{\mathrm{d}\bm{i}_{\mathrm{g}}^{\mathrm{gfm}}}{\mathrm{d}t}
&=
\bm{v}_{\mathrm{f}}^{\mathrm{gfm}}
-
\bm{v}_{\mathrm{pcc}}^{\mathrm{gfm}}
-
r_g^{\mathrm{gfm}}\bm{i}_{\mathrm{g}}^{\mathrm{gfm}}
-
\omega_{\mathrm{oc}}^{\mathrm{gfm}}\ell_g^{\mathrm{gfm}}
\bm{J}\bm{i}_{\mathrm{g}}^{\mathrm{gfm}}
\label{eq:gfm_ig_dyn}
\\[0.3em]
\textit{Algebraic Eqs.} \notag \\[0.3em]
p^{\mathrm{f,gfm}}
&=
(\bm{v}_{\mathrm{f}}^{\mathrm{gfm}})^\top
\bm{i}_{\mathrm{g}}^{\mathrm{gfm}}
\label{eq:gfm_pf}
\\
q^{\mathrm{f,gfm}}
&=
(\bm{v}_{\mathrm{f}}^{\mathrm{gfm}})^\top
\bm{J}\,
\bm{i}_{\mathrm{g}}^{\mathrm{gfm}}
\label{eq:gfm_qf}
\\
\omega_{\mathrm{oc}}^{\mathrm{gfm}}
&=
\omega^{\mathrm{ref,gfm}}
+
k_p^{\mathrm{gfm}}
\bigl(
p^{\mathrm{ref,gfm}}-p^{\mathrm{oc,gfm}}
\bigr)
\label{eq:gfm_woc}
\\
v^{\mathrm{oc,gfm}}
&=
v^{\mathrm{ref,gfm}}
+
k_q^{\mathrm{gfm}}
\bigl(
q^{\mathrm{ref,gfm}}-q^{\mathrm{oc,gfm}}
\bigr)
\label{eq:gfm_voc}
\\
\bm{v}_{\mathrm{vi}}^{\mathrm{gfm}}
&=
v^{\mathrm{oc,gfm}}\bm e_2
-
r_v^{\mathrm{gfm}}\bm{i}_{\mathrm{g}}^{\mathrm{gfm}}
-
\omega_{\mathrm{oc}}^{\mathrm{gfm}}\ell_v^{\mathrm{gfm}}
\bm{J}\bm{i}_{\mathrm{g}}^{\mathrm{gfm}}
\label{eq:gfm_vvi}
\\
\bm{i}_{\mathrm{ref}}^{\mathrm{gfm}}
&=
k_p^{v,\mathrm{gfm}}
\bigl(
\bm{v}_{\mathrm{vi}}^{\mathrm{gfm}}
-
\bm{v}_{\mathrm{f}}^{\mathrm{gfm}}
\bigr)
+
k_i^{v,\mathrm{gfm}}\bm{\xi}^{\mathrm{gfm}}
+
\omega_{\mathrm{oc}}^{\mathrm{gfm}}c_f^{\mathrm{gfm}}
\bm{J}\bm{v}_{\mathrm{f}}^{\mathrm{gfm}}
\label{eq:gfm_iref}
\\
\bm{v}_{\mathrm{cv}}^{\mathrm{gfm}}
&=
k_p^{c,\mathrm{gfm}}
\bigl(
\bm{i}_{\mathrm{ref}}^{\mathrm{gfm}}
-
\bm{i}_{\mathrm{cv}}^{\mathrm{gfm}}
\bigr)
+
k_i^{c,\mathrm{gfm}}\bm{\gamma}^{\mathrm{gfm}}
-
\omega_{\mathrm{oc}}^{\mathrm{gfm}}\ell_f^{\mathrm{gfm}}
\bm{J}\bm{i}_{\mathrm{cv}}^{\mathrm{gfm}}
\label{eq:gfm_vcv}
\\
\bm{v}_{\mathrm{pcc}}^{\mathrm{gfm}}
&=
\bm{R}\!\left(\theta_{\mathrm{oc}}^{\mathrm{gfm}}\right)
\bm{v}_{ri}^{\mathrm{pcc}}
\label{eq:gfm_vpcc}
\\
\bm{i}_{ri}^{\mathrm{g,gfm}}
&=
\bm{R}\!\left(\theta_{\mathrm{oc}}^{\mathrm{gfm}}\right)^\top
\bm{i}_{\mathrm{g}}^{\mathrm{gfm}}
\label{eq:gfm_iri}
\end{align}
}
\end{subequations}

\subsubsection{Grid-Following Inverter}

\begin{subequations}
\label{app:eqgfl}
{\allowdisplaybreaks
\begin{align}
\textit{Differential Eqs.} \notag \\[0.3em]
\frac{1}{\omega_b}\frac{\mathrm{d}\theta_{\mathrm{pll}}^{\mathrm{gfl}}}{\mathrm{d}t}
&=
\omega_{\mathrm{pll}}^{\mathrm{gfl}}-\omega_s
\label{eq:gfl_theta}
\\
\frac{1}{\omega_{\mathrm{lp}}^{\mathrm{gfl}}}
\frac{\mathrm{d}v_{q,\mathrm{pll}}^{\mathrm{gfl}}}{\mathrm{d}t}
&=
v_q^{\mathrm{f,gfl}}-v_{q,\mathrm{pll}}^{\mathrm{gfl}}
\label{eq:gfl_vqpll}
\\
\frac{\mathrm{d}\epsilon^{\mathrm{gfl}}}{\mathrm{d}t}
&=
v_{q,\mathrm{pll}}^{\mathrm{gfl}}
\label{eq:gfl_eps}
\\
\frac{\mathrm{d}\sigma_p^{\mathrm{gfl}}}{\mathrm{d}t}
&=
p^{\mathrm{ref,gfl}}-p_m
\label{eq:gfl_sigmap}
\\
\frac{1}{\omega_z^{\mathrm{gfl}}}\frac{\mathrm{d}p_m}{\mathrm{d}t}
&=
(\bm{v}_{\mathrm{f}}^{\mathrm{gfl}})^\top
\bm{i}_{\mathrm{g}}^{\mathrm{gfl}}
-p_m
\label{eq:gfl_pm}
\\
\frac{\mathrm{d}\sigma_q^{\mathrm{gfl}}}{\mathrm{d}t}
&=
q^{\mathrm{ref,gfl}}-q_m
\label{eq:gfl_sigmaq}
\\
\frac{1}{\omega_f^{\mathrm{gfl}}}\frac{\mathrm{d}q_m}{\mathrm{d}t}
&=
(\bm{v}_{\mathrm{f}}^{\mathrm{gfl}})^\top
\bm{J}\,
\bm{i}_{\mathrm{g}}^{\mathrm{gfl}}
-q_m
\label{eq:gfl_qm}
\\
\frac{\mathrm{d}\bm{\gamma}^{\mathrm{gfl}}}{\mathrm{d}t}
&=
\bm{i}_{\mathrm{ref}}^{\mathrm{gfl}}
-
\bm{i}_{\mathrm{cv}}^{\mathrm{gfl}}
\label{eq:gfl_gamma}
\\
\frac{\ell_f^{\mathrm{gfl}}}{\omega_b}
\frac{\mathrm{d}\bm{i}_{\mathrm{cv}}^{\mathrm{gfl}}}{\mathrm{d}t}
&=
\bm{v}_{\mathrm{cv}}^{\mathrm{gfl}}
-
\bm{v}_{\mathrm{f}}^{\mathrm{gfl}}
-
r_f^{\mathrm{gfl}}\bm{i}_{\mathrm{cv}}^{\mathrm{gfl}}
-
\omega_{\mathrm{pll}}^{\mathrm{gfl}}\ell_f^{\mathrm{gfl}}
\bm{J}\bm{i}_{\mathrm{cv}}^{\mathrm{gfl}}
\label{eq:gfl_icv_dyn}
\\
\frac{c_f^{\mathrm{gfl}}}{\omega_b}
\frac{\mathrm{d}\bm{v}_{\mathrm{f}}^{\mathrm{gfl}}}{\mathrm{d}t}
&=
\bm{i}_{\mathrm{cv}}^{\mathrm{gfl}}
-
\bm{i}_{\mathrm{g}}^{\mathrm{gfl}}
-
\omega_{\mathrm{pll}}^{\mathrm{gfl}}c_f^{\mathrm{gfl}}
\bm{J}\bm{v}_{\mathrm{f}}^{\mathrm{gfl}}
\label{eq:gfl_vf_dyn}
\\
\frac{\ell_g^{\mathrm{gfl}}}{\omega_b}
\frac{\mathrm{d}\bm{i}_{\mathrm{g}}^{\mathrm{gfl}}}{\mathrm{d}t}
&=
\bm{v}_{\mathrm{f}}^{\mathrm{gfl}}
-
\bm{v}_{\mathrm{pcc}}^{\mathrm{gfl}}
-
r_g^{\mathrm{gfl}}\bm{i}_{\mathrm{g}}^{\mathrm{gfl}}
-
\omega_{\mathrm{pll}}^{\mathrm{gfl}}\ell_g^{\mathrm{gfl}}
\bm{J}\bm{i}_{\mathrm{g}}^{\mathrm{gfl}}
\label{eq:gfl_ig_dyn}
\\[0.3em]
\textit{Algebraic Eqs.} \notag \\[0.3em]
\omega_{\mathrm{pll}}^{\mathrm{gfl}}
&=
1
+
k_p^{\mathrm{pll,gfl}}v_{q,\mathrm{pll}}^{\mathrm{gfl}}
+
k_i^{\mathrm{pll,gfl}}\epsilon^{\mathrm{gfl}}
\label{eq:gfl_wpll}
\\
\bm{v}_{\mathrm{pcc}}^{\mathrm{gfl}}
&=
\bm{R}\!\left(\theta_{\mathrm{pll}}^{\mathrm{gfl}}\right)
\bm{v}_{ri}^{\mathrm{pcc}}
\label{eq:gfl_vpcc}
\\
\bm{i}_{\mathrm{ref}}^{\mathrm{gfl}}
&=
\begin{bmatrix}
k_p^{q,\mathrm{gfl}}\bigl(q^{\mathrm{ref,gfl}}-q_m\bigr)
+
k_i^{q,\mathrm{gfl}}\sigma_q^{\mathrm{gfl}}
\\[2pt]
k_p^{p,\mathrm{gfl}}\bigl(p^{\mathrm{ref,gfl}}-p_m\bigr)
+
k_i^{p,\mathrm{gfl}}\sigma_p^{\mathrm{gfl}}
\end{bmatrix}
\label{eq:gfl_iref}
\\
\bm{v}_{\mathrm{cv}}^{\mathrm{gfl}}
&=
k_p^{c,\mathrm{gfl}}
\bigl(
\bm{i}_{\mathrm{ref}}^{\mathrm{gfl}}
-
\bm{i}_{\mathrm{cv}}^{\mathrm{gfl}}
\bigr)
+
k_i^{c,\mathrm{gfl}}\bm{\gamma}^{\mathrm{gfl}}
-
\omega_{\mathrm{pll}}^{\mathrm{gfl}}\ell_f^{\mathrm{gfl}}
\bm{J}\bm{i}_{\mathrm{cv}}^{\mathrm{gfl}}
\label{eq:gfl_vcv}
\\
\bm{i}_{ri}^{\mathrm{g,gfl}}
&=
\bm{R}\!\left(\theta_{\mathrm{pll}}^{\mathrm{gfl}}\right)^\top
\bm{i}_{\mathrm{g}}^{\mathrm{gfl}}
\label{eq:gfl_iri}
\end{align}
}
\end{subequations}

\subsection{Parameter Setup}
\renewcommand{\arraystretch}{1.05}
\setlength{\tabcolsep}{4.2pt}

\begin{longtable}{lll}
\caption{Parameters of the Modified  3-Machine 9-Bus System}
\label{tab:all_params} \\
\toprule
\textbf{Symbol} & \textbf{Value} & \textbf{Description} \\
\midrule
\endfirsthead

\multicolumn{3}{c}{\tablename\ \thetable{} -- continued from previous page} \\
\toprule
\textbf{Symbol} & \textbf{Value} & \textbf{Description} \\
\midrule
\endhead

\midrule
\multicolumn{3}{r}{Continued on next page} \\
\endfoot

\bottomrule
\endlastfoot

\multicolumn{3}{l}{\textit{Synchronous Machine, Exciter, and Governor}} \\
\midrule
$\omega_b$   & $2\pi f_b~\mathrm{rad/s}$ & Base angular frequency \\
$\omega_s$   & $1.0~\mathrm{p.u.}$ & Synchronous speed \\
$H$          & $3.0~\mathrm{s}$ & Inertia constant \\
$D$          & $0.0~\mathrm{p.u.}$ & Damping coefficient \\
$x_d$        & $0.1460~\mathrm{p.u.}$ & $d$-axis synchronous reactance \\
$x_q$        & $0.0969~\mathrm{p.u.}$ & $q$-axis synchronous reactance \\
$x_d'$       & $0.0608~\mathrm{p.u.}$ & $d$-axis transient reactance \\
$x_q'$       & $0.0969~\mathrm{p.u.}$ & $q$-axis transient reactance \\
$T_{d0}'$    & $8.96~\mathrm{s}$ & Open-circuit $d$-axis transient time constant \\
$T_{q0}'$    & $0.31~\mathrm{s}$ & Open-circuit $q$-axis transient time constant \\
$k_a$        & $5.0$ & AVR gain \\
$T_a$        & $0.2~\mathrm{s}$ & AVR time constant \\
$k_e$        & $1.0$ & Exciter coefficient \\
$T_e$        & $0.314~\mathrm{s}$ & Exciter time constant \\
$k_f$        & $0.063$ & Exciter washout gain \\
$T_f$        & $0.35~\mathrm{s}$ & Exciter washout time constant \\
$a_e$        & $0.0039$ & Exciter saturation coefficient \\
$b_e$        & $1.555$ & Exciter saturation coefficient \\
$r$          & $0.15$ & Governor droop coefficient \\
$T_{sv}$     & $0.1~\mathrm{s}$ & Governor servo time constant \\
$T_{ch}$     & $0.5~\mathrm{s}$ & Turbine time constant \\

\midrule
\multicolumn{3}{l}{\textit{Grid-Forming Inverter}} \\
\midrule
$\omega_b$                            & $2\pi f_b~\mathrm{rad/s}$ & Base angular frequency \\
$\omega_s$                            & $1.0~\mathrm{p.u.}$ & System synchronous frequency \\
$\ell_f^{\mathrm{gfm}}$               & $0.08~\mathrm{p.u.}$ & Converter-side filter inductance \\
$r_f^{\mathrm{gfm}}$                  & $0.003~\mathrm{p.u.}$ & Converter-side filter resistance \\
$c_f^{\mathrm{gfm}}$                  & $0.074~\mathrm{p.u.}$ & Filter capacitance \\
$\ell_g^{\mathrm{gfm}}$               & $0.2~\mathrm{p.u.}$ & Grid-side inductance \\
$r_g^{\mathrm{gfm}}$                  & $0.01~\mathrm{p.u.}$ & Grid-side resistance \\
$k_p^{\mathrm{gfm}}$                  & $0.02$ & Active-power droop coefficient \\
$\omega_z^{\mathrm{gfm}}$             & $20.0~\mathrm{rad/s}$ & Active-power filter bandwidth \\
$k_q^{\mathrm{gfm}}$                  & $0.05$ & Reactive-power droop coefficient \\
$\omega_f^{\mathrm{gfm}}$             & $50.0~\mathrm{rad/s}$ & Reactive-power filter bandwidth \\
$r_v^{\mathrm{gfm}}$                  & $0.0~\mathrm{p.u.}$ & Virtual resistance \\
$\ell_v^{\mathrm{gfm}}$               & $0.2~\mathrm{p.u.}$ & Virtual inductance \\
$k_p^{v,\mathrm{gfm}}$                & $0.3947$ & Voltage-loop proportional gain \\
$k_i^{v,\mathrm{gfm}}$                & $49.5953$ & Voltage-loop integral gain \\
$k_p^{c,\mathrm{gfm}}$                & $0.3771$ & Current-loop proportional gain \\
$k_i^{c,\mathrm{gfm}}$                & $335.1032$ & Current-loop integral gain \\
$\omega_{cv}^{\mathrm{gfm}}$          & $2\pi\times 80~\mathrm{rad/s}$ & Voltage-loop bandwidth \\
$\omega_{ci}^{\mathrm{gfm}}$          & $2\pi\times 200~\mathrm{rad/s}$ & Current-loop bandwidth \\

\midrule
\multicolumn{3}{l}{\textit{Grid-Following Inverter}} \\
\midrule
$\omega_b$                            & $2\pi f_b~\mathrm{rad/s}$ & Base angular frequency \\
$\omega_s$                            & $1.0~\mathrm{p.u.}$ & System synchronous frequency \\
$\ell_f^{\mathrm{gfl}}$               & $0.08~\mathrm{p.u.}$ & Converter-side filter inductance \\
$r_f^{\mathrm{gfl}}$                  & $0.003~\mathrm{p.u.}$ & Converter-side filter resistance \\
$c_f^{\mathrm{gfl}}$                  & $0.074~\mathrm{p.u.}$ & Filter capacitance \\
$\ell_g^{\mathrm{gfl}}$               & $0.1~\mathrm{p.u.}$ & Grid-side inductance \\
$r_g^{\mathrm{gfl}}$                  & $0.01~\mathrm{p.u.}$ & Grid-side resistance \\
$k_p^{\mathrm{pll,gfl}}$              & $0.05$ & PLL proportional gain \\
$k_i^{\mathrm{pll,gfl}}$              & $1.42$ & PLL integral gain \\
$\omega_{\mathrm{lp}}^{\mathrm{gfl}}$ & $376.99~\mathrm{rad/s}$ & PLL low-pass filter cutoff frequency \\
$k_p^{p,\mathrm{gfl}}$                & $0.01$ & Active-power loop proportional gain \\
$k_i^{p,\mathrm{gfl}}$                & $0.12$ & Active-power loop integral gain \\
$k_p^{q,\mathrm{gfl}}$                & $0.01$ & Reactive-power loop proportional gain \\
$k_i^{q,\mathrm{gfl}}$                & $0.12$ & Reactive-power loop integral gain \\
$\omega_z^{\mathrm{gfl}}$             & $41.47~\mathrm{rad/s}$ & Active-power measurement filter bandwidth \\
$\omega_f^{\mathrm{gfl}}$             & $41.47~\mathrm{rad/s}$ & Reactive-power measurement filter bandwidth \\
$k_p^{c,\mathrm{gfl}}$                & $0.15$ & Current-loop proportional gain \\
$k_i^{c,\mathrm{gfl}}$                & $0.267$ & Current-loop integral gain \\

\end{longtable}

\end{document}